\relax

\documentclass{article}

\usepackage{microtype}
\usepackage{graphicx}
\usepackage{subfigure}
\usepackage{booktabs} 

\usepackage{hyperref}

\usepackage[accepted]{icml2021}

\usepackage{booktabs}       
\usepackage{amsfonts}       
\usepackage{nicefrac}       
\usepackage{microtype}      
\usepackage{enumitem}
\usepackage{adjustbox}
\usepackage{amsmath}
\usepackage{amsthm}
\usepackage{thmtools}
\usepackage{float}
\usepackage{svg}
\usepackage{natbib}
\usepackage{algorithmic}
\usepackage{algorithm}

\newcommand{\DAPO}{$D_{\scriptscriptstyle APO}$}
\newcommand{\DRCT}{$D_{\scriptscriptstyle RCT}$}
\newcommand{\DOSAPO}{$D_{\scriptscriptstyle OSAPO}$}
\newcommand{\DOSRCT}{$D_{\scriptscriptstyle OSRCT}$}

\newtheorem{theorem}{Theorem}
\declaretheoremstyle[%
  spaceabove=-6pt,%
  spacebelow=6pt,%
  headfont=\normalfont\itshape,%
  postheadspace=1em,%
  qed=\qedsymbol%
]{mystyle} 

\icmltitlerunning{How and Why to Use Experimental Data to Evaluate Methods for Observational Causal Inference}

\begin{document}

\twocolumn[
\icmltitle{How and Why to Use Experimental Data \\to Evaluate Methods for Observational Causal Inference}




\begin{icmlauthorlist}
\icmlauthor{Amanda Gentzel}{UMass,Leidos}
\icmlauthor{Purva Pruthi}{UMass}
\icmlauthor{David Jensen}{UMass}
\end{icmlauthorlist}

\icmlaffiliation{UMass}{College of Information and Computer Sciences, University of
Massachusetts, Amherst, United States}

\icmlaffiliation{Leidos}{Leidos, Reston, Virginia, United States}

\icmlsetsymbol{equal}{*}

\icmlcorrespondingauthor{Amanda Gentzel}{Amanda.M.Gentzel@leidos.com}

\icmlkeywords{Machine Learning, ICML}

\vskip 0.3in
]



\printAffiliationsAndNotice{}  

\begin{abstract}
Methods that infer causal dependence from observational data are central to many areas of science, including medicine, economics, and the social sciences.  A variety of theoretical properties of these methods have been proven, but \textit{empirical} evaluation remains a challenge, largely due to the lack of observational data sets for which treatment effect is known.  We describe and analyze \textit{observational sampling from randomized controlled trials} (OSRCT), a method for evaluating causal inference methods using data from randomized controlled trials (RCTs). This method can be used to create constructed observational data sets with corresponding unbiased estimates of treatment effect, substantially increasing the number of data sets available for empirical evaluation of causal inference methods.  We show that, in expectation, OSRCT creates data sets that are equivalent to those produced by randomly sampling from empirical data sets in which all potential outcomes are available.  We then perform a large-scale evaluation of seven causal inference methods over 37 data sets, drawn from RCTs, as well as simulators, real-world computational systems, and observational data sets augmented with a synthetic response variable.  We find notable performance differences when comparing across data from different sources, demonstrating the importance of using data from a variety of sources when evaluating any causal inference method.


\end{abstract}

\section{Introduction}
Researchers in machine learning and statistics have become increasingly interested in methods that estimate causal effects from observational data.  Such interest is understandable, given the centrality of causal questions in fields such as medicine, economics, sociology, and political science \citep{morgan2015counterfactuals}.  Causal inference has also emerged as an important class of methods for improving the explainability and fairness of machine learning systems, since causal models can explicitly represent the underlying mechanisms of systems and their likely behavior under counterfactual conditions \citep{kusner2017counterfactual,pearl2019seven}.

However, evaluating causal inference methods is far more challenging than evaluating purely associational methods.  Both types of methods can be analyzed theoretically.  However, \textit{empirical} analysis---long a driver of research progress in machine learning and statistics---has been increasingly recognized as vital for research progress in causal inference \citep[e.g.,][]{Dorie2019,Gentzel2019}, and empirical evaluation is substantially more challenging to perform in the case of causal inference.  Many associational models (e.g., classifiers and conditional probability estimators) can be evaluated using cross-validation or held-out test sets.  However, causal inference aims to estimate the value or distribution of an outcome variable \textit{under intervention}, and evaluating such estimates requires an alternative route to estimating the effects of such interventions.



Most available data sets are either experimental (which can yield unbiased estimates of treatment effect) or observational (for which treatment effect is unknown).  Since most causal inference methods are designed to infer causal dependence from observational data, accurate evaluation requires both observational data and corresponding unbiased estimates of treatment effect.  Several recent efforts have attempted to address this problem \citep[e.g.,][]{Dorie2019, Gentzel2019, Tu2019, Shimoni2018}, most of which collect or modify data specifically for the purpose of evaluation.  Some approaches induce dependence between variables in specially constructed or selected data, while others re-purpose a simulator to produce data for evaluation.  These approaches are promising and beneficial to the community, but creating individual, specialized new data sets is difficult and time-consuming, limiting the number of data sets available.  To this point, most data sets that are easy to collect (such as synthetic data and simulators) are generally unrealistic, while data sets with a high degree of realism require significant effort to obtain.

We argue for exploiting a largely untapped source of data for evaluating causal inference methods: randomized controlled trials (RCTs).  RCTs are designed and conducted for the express purpose of providing unbiased estimates of treatment effect.  Many RCT data sets are publicly available, and that number is only increasing, allowing for the collection of a large number of realistic data sets.  Previous work has described how to non-randomly sample a specialized type of experimental data (one in which all potential outcomes are observed) to create \textit{constructed observational data sets}.\footnote{The term ``constructed observational data" denotes empirical data to which additional properties common in observational data (e.g., confounding) have been synthetically introduced.  
This term is distinct from \textit{constructed observational studies}, which are studies that collect and compare both experimental and observational data from the same domain (see Section \ref{sec:related-work}).}  Surprisingly, this basic approach can be modified to produce constructed observational data from RCTs as well. 

The core idea of non-random sampling of data from RCTs is not original to this paper.  This basic approach has been used sporadically for at least a decade to evaluate algorithms for observational causal inference \cite{hill2011bayesian}. However, these uses have typically been described only in passing and have often been limited to a single data set.
This basic approach has also been used more widely to evaluate algorithms for learning policies for contextual bandits \cite{li2011unbiased}. However, this work is almost unknown within the community of researchers studying observational causal inference. The properties and wide utility of this approach are sufficiently unknown that a more general analysis and discussion is warranted. Section \ref{sec:related-work} provides additional discussion of related work.

We use data from multiple RCTs and also from a wide variety of other data sources to perform a large-scale evaluation of several causal inference methods. These data sources can be broadly categorized as RCTs, simulators \cite{Nemo, Tu2019, miller2020whynot}, real-world computational systems \cite{Gentzel2019} and observational data sets augmented with a synthetic response variable \cite{Dorie2019, Shimoni2018}. This type of large-scale evaluation from a diverse set of data sources provides us with an unprecedented opportunity to analyze systematic differences in the performance of the methods not only due to characteristics of the algorithms but also due to differences in data source. 

Specifically, we: 
(1) Propose that a known, but sporadically used, method for inducing confounding bias in RCT data become part of the standard evaluation suite for causal inference methods;
(2) Prove that this approach is equivalent, in expectation, to the data generating process assumed by the potential-outcomes framework, a longstanding theoretical framework for causal inference;
(3) Demonstrate the feasibility of this approach by applying multiple causal inference methods to observational data constructed from RCTs; and 
(4) Perform a large-scale evaluation of seven causal inference methods on 37 data sets drawn from multiple empirical data sources.\footnote{Pointers to the data sets used in this paper, and R code to perform observational sampling, are available at \url{https://github.com/KDL-umass/papers/tree/main/causal-eval-rct-icml-2021}.}


\section{Creating Observational Data from Randomized Controlled Trials}


Consider a data generating process that produces a binary treatment $T \in \{0,1\}$, outcome $Y$, and multiple covariates $C=\{C_1,C_2,...C_k\}$, each of which may be causal for outcome.\footnote{For ease of exposition, we describe the approach using binary treatment, but the approach is more general.} We define $Y_i(t)$ to be the outcome for unit $i$ under treatment $t$, referred to as a \textit{potential outcome}.  For each unit $i$, both treatment values $T_i=0$ and $T_i=1$ are set by intervention and both potential outcomes $Y_i(1)$ and $Y_i(0)$ are measured.  We refer to this type of data, where all potential outcomes are observed, as \textit{all potential outcomes} (APO) data, denoted \DAPO.  Note that, due to the use of explicit interventions, such a data generating process produces \textit{experimental}, rather than observational, data.  

Recently, some researchers \citep{louizos2017causal, Gentzel2019} have proposed sampling from APO data to produce constructed observational data.  Such data sets are produced by probabilistically sampling a treatment value (and its corresponding outcome value) for every unit based on the values of one or more covariates. We refer to $C^b \subseteq C$ as the \textit{biasing covariates} and selected treatment random variable as $T^{s}$, thus $P(T^s|C^b) := f(C^b)$. This procedure, shown in Algorithm \ref{alg:APO-biasing-algorithm}, induces causal dependence between $C^b$ and $T^s$, creating a confounder when $C^b$ also causes $Y$.  We refer to such a data generating process as \textit{observational sampling from all potential outcomes} (OSAPO) and denote a given data set generated in this way as \DOSAPO.  OSAPO is the data generating process assumed under the potential outcomes framework \citep{rubin2005causal}.  

Data sets produced by OSAPO are extremely useful for evaluating causal inference methods.  Causal inference methods can estimate treatment effect in \DOSAPO, and these estimates can be compared to estimates derived from \DAPO.  Furthermore, the process of inducing bias by sub-sampling allows for a degree of control that can be exploited to evaluate a method's resilience to confounding, by systematically varying the strength and form of dependence and whether variables in $C^b$ are observed.  However, very few experimental data sets exist that record all potential outcomes for every unit, severely limiting the applicability of this approach.   

\subsection{Observational Sampling of RCTs}

Now consider a slightly different data generating process, in which treatment is randomly assigned and only one potential outcome is measured for each unit $i$, producing either $Y_i(1)$ or $Y_i(0)$, but not both.  This is the data generating process typically implemented by RCTs, in which every unit is randomly assigned a single treatment value, and the outcome for that treatment is measured.  Vast numbers of RCTs are conducted each year, and data sets from many of them are available publicly.  In addition, growing efforts toward open science are continually increasing the number of publicly available RCT data sets.

This raises an intriguing research question: \textit{Can RCTs be sub-sampled to produce constructed observational data sets with the same properties as those produced by APO sampling?}

\begin{figure*}[ttt!]
\small
\begin{minipage}[t]{0.5\textwidth}
\vspace{0pt}
\begin{algorithm}[H]
\begin{algorithmic}
    \STATE {\bfseries Input:} APO data set \DAPO, biasing covariates $C^b$
    \STATE {\bfseries Output:} Biased data set \DOSAPO
    \FORALL{units $i \in D$}
        \STATE $p \leftarrow f(C^b_i)$
        \STATE $t_s \leftarrow Bernoulli(p)$
        \STATE $o \leftarrow$ row in $D_{\scriptscriptstyle APO}$ corresponding to $(i,t_s)$
        \STATE $D_{\scriptscriptstyle OSAPO} \leftarrow D_{\scriptscriptstyle OSAPO} \cup o$
    \ENDFOR
    \STATE {\bfseries Return} \DOSAPO
\caption{Observational sampling from \newline all potential outcomes (OSAPO)}\label{alg:APO-biasing-algorithm}
\end{algorithmic}
\end{algorithm}
\end{minipage}
\hspace{.05cm}
\begin{minipage}[t]{0.5\textwidth}
\vspace{0pt}
\begin{algorithm}[H]
\begin{algorithmic}
    \STATE {\bfseries Input:} RCT data set \DRCT, biasing covariates $C^b$
    \STATE {\bfseries Output:} Biased data set \DOSRCT
    \FORALL {unit $i \in D$}
        \STATE $p \leftarrow f(C^b_i)$
        \STATE $t_s \leftarrow Bernoulli(p)$
        \IF {$(i,t_s) \in D$}
            \STATE $o \leftarrow$ row in $D_{\scriptscriptstyle RCT}$ corresponding to $(i,t_s)$
            \STATE $D_{\scriptscriptstyle OSRCT} \leftarrow D_{\scriptscriptstyle OSRCT} \cup o$
        \ENDIF
    \ENDFOR
    \STATE {\bfseries Return} \DOSRCT
\caption{Observational sampling from \newline randomized controlled trials (OSRCT)}\label{alg:RCT-biasing-algorithm}
\end{algorithmic}
\end{algorithm}
\end{minipage}
\caption{Two procedures for sampling constructed observational data sets from experimental data. \textit{Left}: From all potential outcomes (APO) data. \textit{Right}: From randomized controlled trial (RCT) data.  For some function $f:\mathcal{D}(C^b) \rightarrow \{x \in \mathbb{R}:0<x<1\}$}
\end{figure*}

\begin{figure*}
\centering
\includegraphics[width=.75\textwidth]{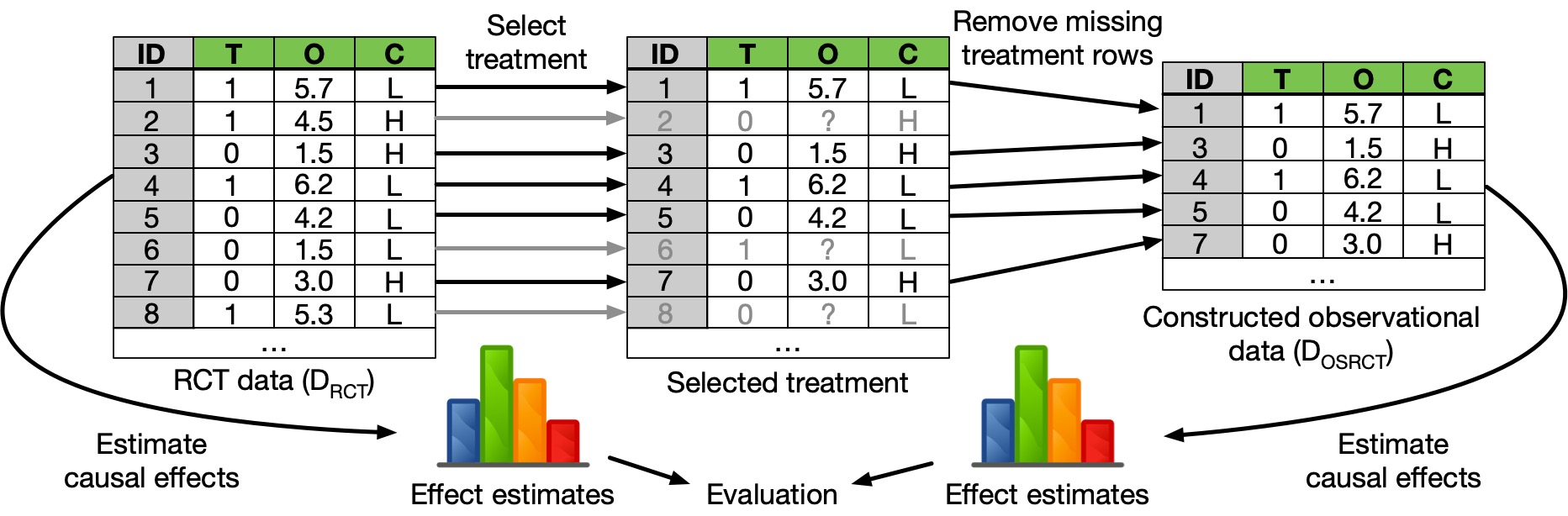}
\caption{The process of creating observational-style data from a randomized controlled trial.}
\label{fig:RCT-biasing-protocol}
\end{figure*}

We describe one such sampling procedure in Algorithm \ref{alg:RCT-biasing-algorithm} --- \textit{observational sampling from randomized controlled trials} (OSRCT)---which produces a data sample denoted \DOSRCT.  As in APO sampling, covariates $C^b$ bias the selection of a single treatment value, $t_s$, for every unit $i$.  If unit $i$ actually received the selected treatment $t_s$, we add $i$ to \DOSRCT.  Otherwise, that unit is ignored.  As we show below in Theorem \ref{thm:half-size}, when treatment is binary and the treatment and control groups are equal in size, the resulting constructed observational data set is, in expectation, half the size of the original, regardless of the form of the biasing.  As discussed in Section \ref{sec:evaluation-methods}, a causal inference method can then be applied to this data, and the results can be compared to the unbiased effect estimate from the original RCT data.  This basic approach is shown in Figure \ref{fig:RCT-biasing-protocol}.

An RCT can be thought of as a data set where one potential outcome for every unit is missing at random.  Since OSRCT uses the biasing covariates to select treatment, and treatment was assigned randomly, the sub-sampling process only changes the dependence between the biasing covariates and treatment.  This is the same as in APO sampling.  The probability of a given unit-treatment pair being included in the sub-sample is proportional in APO and RCT sampling. That is, \DOSRCT\ is equivalent to a random sample of \DOSAPO.  We demonstrate this empirically, using an APO data set converted into an RCT data set, and provide a proof, in the supplementary material.

\begin{theorem}
For RCT data set $D_{\scriptscriptstyle RCT}$, APO data set $D_{\scriptscriptstyle APO}$, and binary treatment $T \in \{0,1\}$ with $P(T=1) = P(T=0) = 0.5$ in $D_{\scriptscriptstyle RCT}$, and units $i$, $P_{D_{OSRCT}}(T_i=t) = 0.5*P_{D_{OSAPO}}(T_i=t)$, for all units $i$ and treatment values $t$.
\label{thm:equivalent-to-OSAPO}
\end{theorem}



Note that while Theorem \ref{thm:equivalent-to-OSAPO} assumes equal probability of treatment and control, the approach generally applies even when $P(T=1) \neq 0.5$.  In this case, instead of sub-sampling $D_{\scriptscriptstyle OSAPO}$ by a factor of 0.5, the scaling factor is selected based on the treatment value.  Since treatment is based solely on the value of the biasing covariates, this is equivalent to modifying the form of the biasing function.

One potential disadvantage of this approach is that sub-sampling to induce bias necessarily reduces the size of the resulting sample.  Somewhat surprisingly, however, the degree of this reduction does not depend on the intensity of the biasing.

\begin{theorem}
For binary treatment $T \in \{0,1\}$ and RCT data set \DRCT, if either $P(T=1)=P(T=0)=0.5$, or $E[P(T^s=1|C^b)] = 0.5$, then $E[|D_{\scriptscriptstyle OSRCT}|] = 0.5|D_{\scriptscriptstyle RCT}|$. 
\label{thm:half-size}
\end{theorem}
A proof is provided in the Supplementary Material.

One downside of OSRCT is that it reduces the sample size available for causal inference.  One alternative is to \textit{reweight} units, rather than sub-sample them, according to $P(T^{s}_{i}=t_i|C^b_i)$.  This has the benefit of basing causal estimates on every unit in the RCT rather than only those in the sub-sample.  This approach requires that the causal inference method under evaluation accepts unit-level weights. However, when this is an option, using weighting rather than sub-sampling can be a useful alternative when RCTs with small sample size \footnote{Thanks to an anonymous reviewer for the recommendation of this approach}. We compared the estimates obtained by weighting to the estimates obtained by sub-sampling for causal inference methods that accept unit-level weights and found the results to be similar. Comparison results are provided in the supplementary material.

\subsection{What Can OSRCT Evaluate?}
\label{sec:evaluation-methods}
The constructed observational data created by OSRCT has a substantial benefit over purely observational data: Unbiased estimates of causal effect can be obtained from the original RCT data, which can be compared to effect estimates from causal inference methods.  A well-designed RCT enables the unbiased estimation of the sample average treatment effect (ATE) as $E[y_i(1)] - E[y_i(0)]$ = $E[y_i|t_i = 1] - E[y_i|t_i=0]$, where $t_i$ denotes the actual treatment received by unit $i$.  This estimate can be compared to estimates made by causal inference methods applied to the constructed observational data.

Unlike APO data, RCT data only contains one treatment-outcome pair for every unit, limiting both the available effect estimates and how these data sets can be used.  RCTs measure the effect of a single randomized intervention $do(T_i=t_i)$ for every unit in the data set.  Thus, we cannot estimate individual treatment effect (ITE) from RCT data, a measurement that is available when using APO data.

However, while RCTs do not measure individual-level treatment effects, they do measure individual-level outcomes under intervention. Thus, OSRCT data \textit{can} be used to evaluate a method's ability to estimate unit-level outcome under intervention.  Any causal inference method that can estimate $E[Y|do(T=t)]$ can be evaluated by comparing those estimates against measurements in the RCT data.  In addition, the data discarded during biased sub-sampling in OSRCT can be treated as a held out test set, where the dependence between the biasing covariates and treatment is the complement of that in \DOSRCT.  We call the data discarded during OSRCT the ``complementary sample''.  Because we know the functional form of the dependence between the biasing covariates and treatment, we can weight the data points in the complementary sample  according to their probability of being included in the accepted sample.  In aggregate, this type of weighting allows the complementary sample to approximate the distribution of the training data, and thus be used for testing.  This is equivalent to inverse propensity score weighting \citep{rosenbaum1983central}. The supplementary material contains more details on using the complementary sample for evaluation, including a proof.

\subsection{Assumptions, Limitations, and Opportunities}
The validity of evaluation with OSRCT depends on several standard assumptions about the validity of the original RCT.  Specifically, it assumes that treatment assignment is randomized and that all sampled units complete the study (no ``drop-out''). Intriguingly, one standard assumption---that intent to treat does not differ from actual treatment---is not necessary.  Even if this assumption is violated, the estimated treatment effect will correspond to the effect of intending to treat, and this estimand can still be used to evaluate the effectiveness of methods for observational causal inference.

Evaluation with OSRCT has some limitations.  OSRCT can induce dependence between any covariate and treatment, but not between any covariate and outcome.  In addition, while the original RCT data can yield an unbiased estimate of the effect of treatment on outcome, it cannot produce such estimates for any other pair of variables.

Constructing observational data also provides some unique opportunities.  OSRCT produces data with non-random treatment assignment, and allows for variation in the level and form of that non-randomness.  Additional features of observational studies can also be simulated, such as measurement error, selection bias, positivity violations, and hidden confounding (by hiding one of the biasing covariates).  While some of these may further reduce the sample size of the constructed observational data due to additional sub-sampling, this process of constructing observational data can allow for the evaluation of a causal inference method's robustness to many features of real-world data.

One potential concern about the use of RCTs is their realism.  While RCTs are empirical, the situations in which RCTs are conducted may differ from many observational settings.  For example, RCTs are performed in settings where treatment can be randomized, where relevant covariates can be collected, where outcome can be easily measured, and, in many cases, where there is an assumption that a causal effect is likely.

For the purpose of evaluation, though, the important question is not whether the causal effect estimates from the RCT data are representative of the population of interest, but whether the performance of causal inference methods on samples from the RCT data are representative of their performance on the population of interest.  While this is not guaranteed, this is a far weaker assumption than requiring that the RCT population match the desired observational population.

\section{Related Work}
\label{sec:related-work}

OSRCT has been used for at least a decade in two different settings. The first is the most similar to what we describe in this paper, in which OSRCT has been applied sporadically to evaluate methods for causal inference from observational data. An early use is Hill \citeyearpar{hill2011bayesian}, in which the author non-randomly samples data from an RCT---the Infant Health and Development Program (IHDP)---to produce a single observational-style data set for evaluating Bayesian additive regression trees (BART). The resulting constructed observational data set has subsequently been reused by others to evaluate a variety of methods for observational causal inference \citep[e.g.,][]{shalit2017estimating,atan2018deep,yao2018representation,saito2020counterfactual}.  Less frequently, researchers have applied OSRCT or closely related approaches to other data sets drawn from RCTs or natural experiments \citep[e.g.,][]{arceneaux2006comparing,kallus2018removing,kallus2018confounding,witty2020causal,zhang2021bounding}.

Despite these sporadic uses, observational sampling from RCTs is not widely known or used within the causal inference community.  For example, two recent papers that systematically review existing evaluations of causal inference methods---Dorie et al. (\citeyear{Dorie2019}) and Gentzel et al. (\citeyear{Gentzel2019})---do not even mention this approach, despite the fact that it overcomes many of the most serious threats to validity for evaluation studies (e.g., reproducibility, realistic data distributions and complexity of treatment effects, multiple possible levels of confounding). It has not been explicitly formalized nor have its advantages been clearly described.  As a result, it is rarely used and it has never been systematically compared to alternative approaches to evaluation.

The second setting in which OSRCT has been applied is off-line evaluation of contextual bandit policies \citep{li2011unbiased}.  Specifically, Li et al. show how to evaluate a (non-random) contextual bandit policy by sampling from the data produced by a randomized policy.
This method is widely employed to evaluate methods in fields such as computational advertising and recommender systems \citep[e.g.,][]{tang2013automatic, tang2014ensemble, zeng2016online}, and it has been extended with approaches such as bootstrapping \citep{mary2014improving}.

Our use of OSRCT in this paper exploits the same idea but in a subtly different setting.  In our setting, we have no interest in estimating the effect of a contextual policy that is known to the agent.  Instead, our goal is to determine how well a given method estimates the average or conditional treatment effect (which, in contextual bandits, would be formulated as the reward difference between two specific policies), even though the algorithm only has access to the actions and outcomes of a single unknown and non-randomized policy.

In addition to these two settings that have directly used OSRCT, other prior work has explicitly focused on empirical evaluation methods for observational causal inference.  The ideal method would score highly on at least three characteristics: \textit{data availability} (many data sets with the required characteristics can be easily obtained to avoid overgeneralizing from a small sample); \textit{internal validity} (differences between estimated treatment effect and the standard can only be attributed to bias in the estimator rather than biases due to how data sets were created); and \textit{external validity} (the performance of the estimator will generalize well to other settings).  Of three broad classes of prior work, each suffers from deficiencies and none clearly dominate.

Some prior work uses \textit{observational data sets with known treatment effect}.  One approach gathers observational data about phenomena that are so well-understood that the causal effect is obvious \citep[e.g.,][]{mooij2016distinguishing}, but such situations are rare, limiting data availability.  Another approach uses data from matched pairs of observational and experimental studies \citep[e.g.,][]{Dixit2016, sachs2005causal,jaciw2016assessing}.  Such data sets appear to represent a nearly ideal scenario for evaluating methods for inferring causal effect from observational data, but pairs of directly comparable observational and experimental studies have low data availability and using paired studies with different settings or variable definitions can greatly reduce internal validity.  Some ``constructed observational studies'' intentionally create matched pairs of experimental and observational data sets \citep[e.g.,][]{lalonde1986evaluating, Hill2004, Shadish2008can}, but these studies also have low data availability.

Another class of prior work generates observational data from \textit{synthetic or highly controlled causal systems} \citep[e.g.,][]{Tu2019,Gentzel2019,louizos2017causal,Kallus2018,athey2021using}.  In this way, the treatment effect is either directly known or can be easily derived from experimentation.  Observational data is typically obtained via some biased sampling of the experimental data, often a variety of APO sampling.  In the case of entirely synthetic data, data availability and internal validity are both high, but external validity is low, and such studies are often criticized as little more than demonstrations.  External validity typically increases somewhat for highly controlled causal systems, but data availability drops significantly.

The final and newest class of existing work augments \textit{an existing observational study with a synthetic outcome}, replacing the original outcome measurement with the result of a synthetic function \citep[e.g.,][]{Dorie2019,Shimoni2018}.  Given the synthetic nature of the outcome function, the causal effect is known.  This class of approach has relatively high data availability, and it trades some loss of external validity (because real outcome measurements are replaced with synthetic ones) to gain internal validity (because the true treatment effect is known).  Note particularly that \textit{both} the treatment effect \textit{and} the confounding are synthetic, because the function that determines the synthetic outcome determines how both the treatment and potential confounders affect the value of outcome.

The approach proposed here—OSRCT—augments, rather than replaces, existing approaches. It occupies a unique position because it simultaneously has fairly high data availability, internal validity, and external validity.  OSRCT's data availability is relatively high because it can use data from any moderately large RCT.  Only synthetic data generators and synthetic outcome approaches have higher data availability, but both suffer in terms of external validity.  OSRCT's internal validity is relatively high because there exist many well-designed RCTs.  Using synthetic data generators or highly controlled causal systems will typically produce somewhat higher internal validity, as will observational data with synthetic outcomes, but this is done at the cost of external validity or data availability.  Finally, OSRCT's external validity is relatively high because the distributions of all variables and the outcome function occur naturally, while only the confounding is synthetic. Only observational studies with known treatment effect have higher external validity, and these suffer from severe limitations on data availability.

\section{Are RCT Data Sets Available?}
OSRCT has the benefit of leveraging existing empirical data rather than requiring the creation of new data sets specifically for evaluating causal inference methods, but it does require that data from RCTs be available and generally accessible to causality researchers.  Fortunately, this is increasingly the case.  While many repositories that host RCTs are restricted for reasons of privacy and security, many other repositories allow access with only minimal restrictions.  In some cases, access requires only registering with the repository and agreeing not to re-distribute the data or attempt to de-anonymize it.  As long as these data sharing agreements are adhered to, such data can be easily acquired by causality researchers.  An even larger set of repositories restricts access but will make data available upon reasonable request.  Additional information about some such repositories can be found in the supplementary material.

In addition, funding agencies and journals are increasingly requiring that researchers make anonymized individual patient data available upon reasonable request \citep{Godleee7888, Ohmanne018647}.  For example, the United States' National Institutes of Health (NIH) recently requested public feedback on a proposed data sharing policy with the aim of improving data management and the sharing of data created by NIH-funded projects \citep*{NIH2019}.  There is also increasing awareness and discussion in the medical community about the importance of sharing individual patient data, to allow for greater transparency and re-analysis \citep{Drazen2015, kuntz2019individual, Banzi2019, Suvarna2015}. All of this suggests increasing availability of individual patient data from randomized controlled trials.


\section{Evaluation}

To assess how well RCT data sets work for evaluation, we performed a large-scale evaluation using RCT data, as well as data from three additional sources: computational systems, synthetic-response data, and simulators.  

\subsection{Data}

\textbf{Computational Systems.}
We use the computational system data sets provided by \citet{Gentzel2019}. In these data sets, all potential outcomes are observed, so we refer to these as APO data sets.  These data sets are collected from three computational systems: queries executed by a Postgres database, HTTP requests executed by web servers on the open internet, and programs compiled under the Java Development Kit.  For each data set, we selected a single treatment-outcome pair and a biasing covariate, matching the setup in \citet{Gentzel2019}.

\textbf{Synthetic-Response Data.}
Many data sets for evaluation were created for the ACIC Competition \cite{Dorie2019} and the IBM Causal Inference Benchmarking Framework \cite{Shimoni2018}.  These data sets were created using a set of real-world covariates and then simulating both a treatment and an outcome.  As in APO and RCT data, treatment is selected synthetically based on values of one or more covariates. However, the outcome (sometimes referred to as the `response surface') is also generated synthetically.  Both the ACIC Competition and the IBM Causal Inference Benchmarking Framework created a large number of data sets, with varying treatment and outcome functions.  We selected five data sets from each, for a total of ten data sets, to use for our evaluation.

\textbf{Simulators.}
We used data sets from three simulators of varying complexity: a simulator of neuropathic pain \cite{Tu2019}; Nemo \cite{Nemo}, a simulator of population dynamics; and three simple simulators from the WhyNot Python package \cite{miller2020whynot}.  For both Nemo and the neuropathic pain simulator, we chose three distinct treatment-outcome pairs, generating three data sets for each.  For the WhyNot simulators, we chose three separate simulators and generated a single data set from each, resulting in nine total data sets from simulators. 

\begin{figure*}
\centering
\includegraphics[width=.9\textwidth]{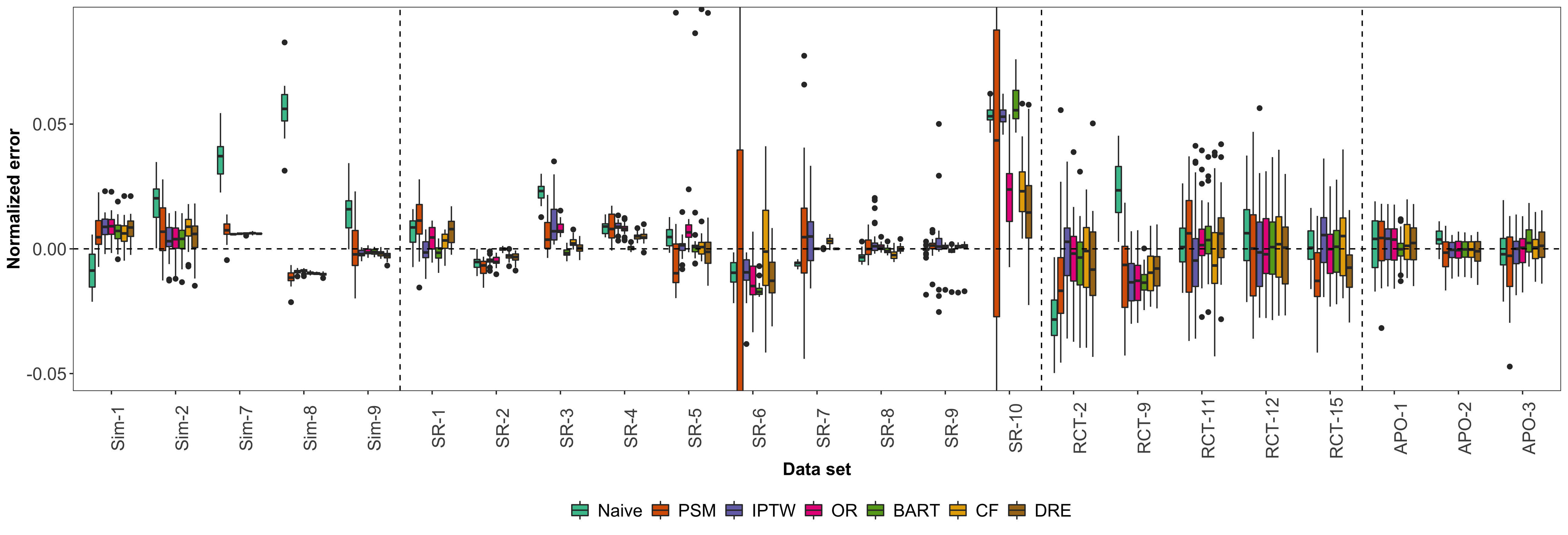}
\caption{Normalized error in estimating ATE for data sets with continuous outcome. Sim denotes simulator, SR denotes synthetic-response data sets, RCT denotes randomized controlled trials and APO denotes computational systems.}
\label{fig:2000_sample_ATE}
\end{figure*}

\textbf{Randomized Controlled Trials.}
We selected data sets from six repositories: Dryad \citeyearpar{Dryad}, the Yale Institution for Social and Policy Studies Repository \citeyearpar{ISPS}, the NIH National Institute on Drug Abuse Data Share Website \citeyearpar{NIDA}, the University of Michigan's ICPSR repository \citeyearpar{ICPSR}, the UK Data Service \citeyearpar{UKDataService}, and the Knowledge Network for Biocomplexity \citeyearpar{KNB}.  These repositories were selected because they contained RCT data, were reasonably well-documented, and had a simple data access process.  None of these repositories house RCT data exclusively, so some search and filtering was necessary to identify relevant data sets.  We also used a stand-alone RCT from a study of artificial cultural markets \cite{salganik2006experimental}.

These four sources of data represent a range of realism.  Of the simulators, the WhyNot simulators are the most simplistic and, by design, are not intended to be realistic representations of the world.  The Nemo and neuropathic pain simulators are a bit more realistic, since they were designed by experts within their respective communities with the intent of accurately modeling their respective systems.  The synthetic-response data sets are even more realistic, because the covariate distribution is from an empirical source.  However, both the treatment and outcome are synthetic.  The APO and RCT data sets have realistic covariate distributions and response functions, with synthetic treatment functions.  While the APO data sets are limited to a narrow set of domains, the RCT data sets come from a wide range of studies in different fields.  This spectrum allows us to assess how algorithm performance differs under different levels of realism.  Further details about all of the data sets are included in the supplementary material.

\subsection{Algorithms}
Due to the nature of the ground truth in the selected data sets (treatment effect of a single treatment on a single outcome), we focused our evaluation on causal inference methods that estimate average treatment effect.  We chose seven methods to evaluate:
propensity score matching (PSM), inverse probability of treatment weighting (IPTW) \cite{rosenbaum1983central}, outcome regression (OR), BART \cite{chipman2007bayesian}, causal forests (CF) \cite{Wager2017}, doubly-robust estimation (DRE) \cite{funk2011doubly}, and a neural network-based method (NN) \cite{shi2019adapting}.  As a baseline for comparison, we also included a naive method that simply estimates $E[Y|T=1] - E[Y|T=0]$ from the observational data. Details about these methods can be found in the supplementary material.

\subsection{Experiments}
For the RCT, APO and simulator data sets\footnote{The synthetic-response data sets were, by design, pre-biased, so no additional processing was necessary}, we selected a single treatment and outcome pair, and selected a biasing covariate that was pre-treatment and correlated with outcome.  For each data set, we calculated the unbiased ATE to use as ground truth.  We then sub-sampled, as shown in Algorithms \ref{alg:APO-biasing-algorithm} and \ref{alg:RCT-biasing-algorithm}, to produced a biased data set.
All algorithms were applied to the biased data, producing estimates of ATE.
This process was repeated for 30 trials.  For data sets with more than 2,000 individuals, we sub-sampled to 2,000, to keep sample sizes comparable between data sets.

For data sets with a binary outcome, we used risk difference instead of ATE, calculated as $P(Y=1|do(T=1)) - P(Y=1|do(T=0))$.  Error in ATE and risk difference estimation was calculated as the absolute difference between the predicted value and the ground truth value calculated on the unbiased data.  ATE and outcome estimates were normalized by the range of the outcome variable for easier comparison.

It is important to note that, while the biasing procedure was the same for all the RCT, Simulator, and APO data sets, the strength of the bias may differ substantially.  This is because two factors influence the strength of the confounding bias: (1) the strength of the treatment sampling function $f(C^b)$; and (2) the strength of the latent dependence between the biasing covariates and outcome.  In order to ensure that confounding bias was introduced, we only chose biasing covariates that were correlated with outcome.  However, the strength and nature of this dependence can vary significantly.  This means that even if we induce strong dependence in the sampling procedure, the overall confounding bias may be weak, due to a weak effect of the biasing covariates on outcome.  To assess how much confounding bias is introduced in each data set, the naive algorithm (described above) serves as a simple estimate of that bias, by showing the error induced by not accounting for the biasing covariates.

\begin{figure}[h]
\centering
\includegraphics[width=.45\textwidth]{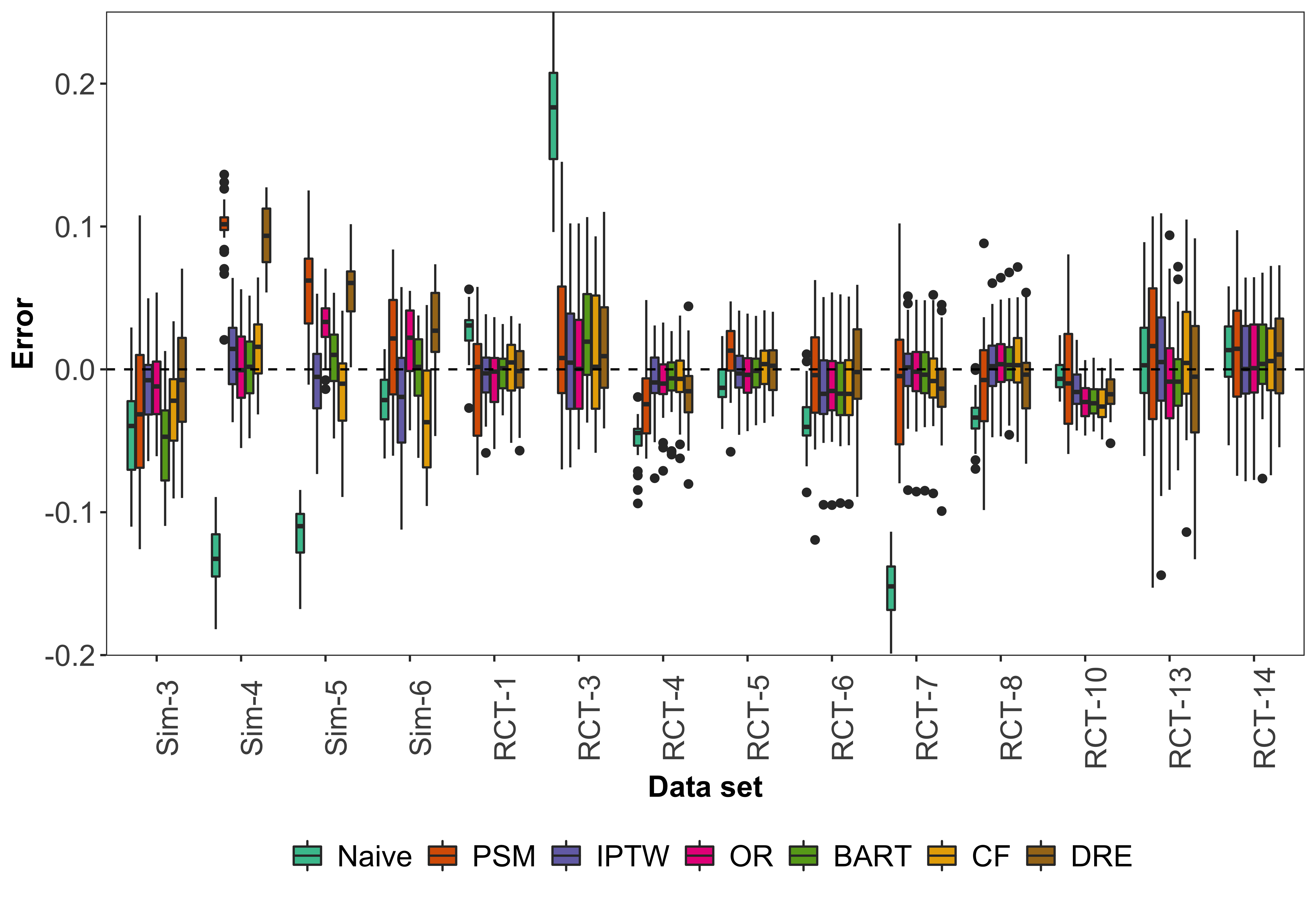}
\caption{Error in estimating risk difference for data sets with binary outcomes.}
\label{fig:2000_sample_rd}
\end{figure} 

\begin{figure*}[h]
\centering
\begin{minipage}[b]{.45\textwidth}
\includegraphics[width=\textwidth]{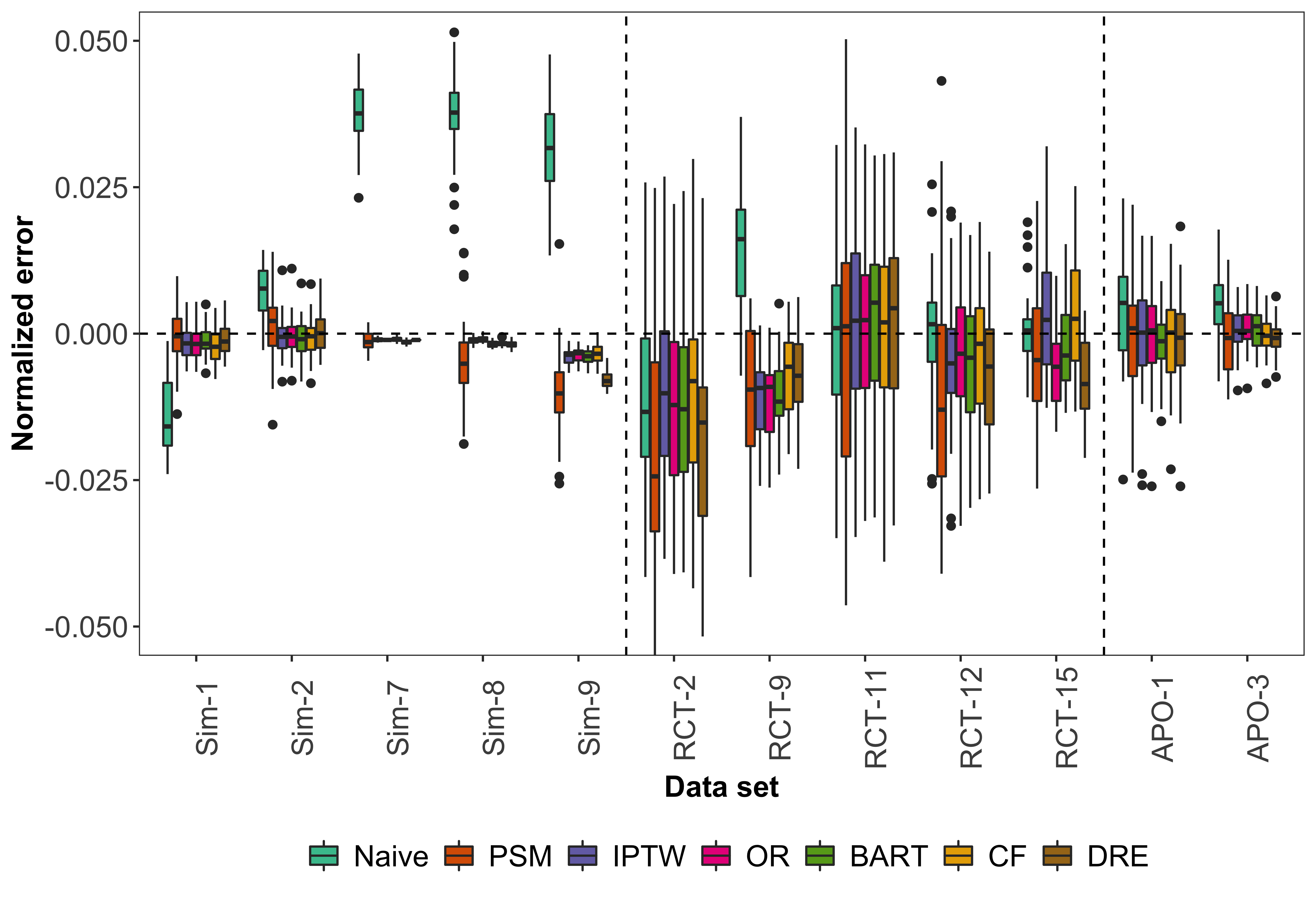}
\caption{Normalized error in estimating ATE with two biasing covariates, for data sets with continuous outcome}\label{fig:two_bias_ATE}
\end{minipage}\qquad
\begin{minipage}[b]{.45\textwidth}
\includegraphics[width=\textwidth]{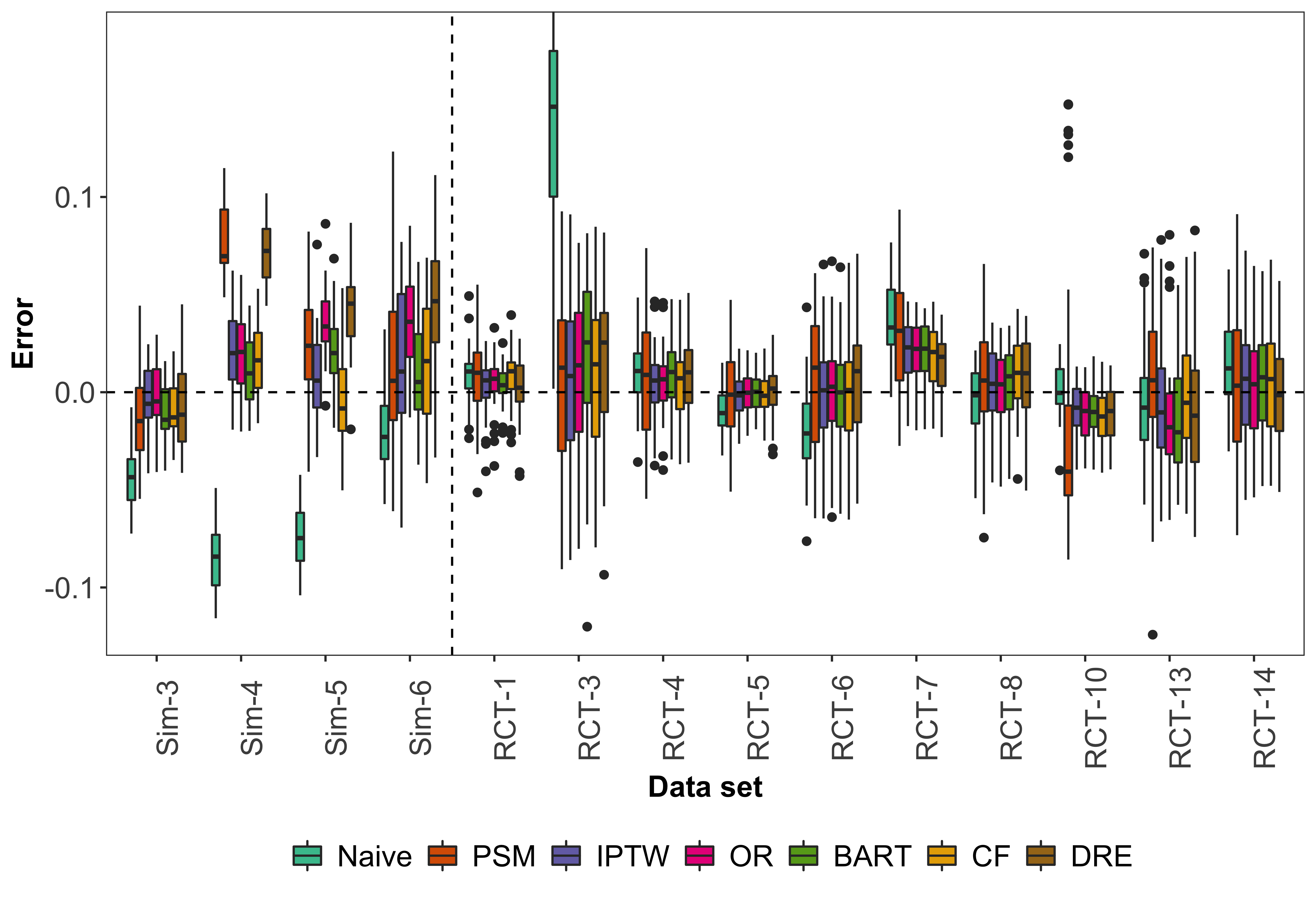}
\caption{Error in estimating risk difference with two biasing covariates, for data sets with binary outcome}\label{fig:two_bias_rd}
\end{minipage}
\end{figure*}

\subsection{Results}
Effect estimation results for the experimental setup described above are shown in Figure \ref{fig:2000_sample_ATE} for data sets with continuous outcomes and Figure \ref{fig:2000_sample_rd} for data sets with binary outcomes.  These results can be analyzed in two main ways: comparing the performance of different algorithms within each data set, and comparing how algorithm performance differs between different data sets.  Most evaluations in the literature focus solely on comparing performance within individual data sets.  Evaluating across data sources, though, can also provide a useful understanding of how different data design choices affect estimates of relative performance.

All methods perform well overall, with error typically centered around zero. Propensity score matching consistently has the highest variability.  This is consistent with the literature, which shows that the pruning done by propensity score matching can increase data set imbalance, and thus increase estimation bias, when matching solely on the propensity score \cite{king2019propensity}.


While performance between the methods is very similar for most for the RCT data sets, the synthetic-response data sets show substantially more variability between different algorithms.  For example, IPTW has higher variability on some synthetic-response data sets, BART performs well overall, and BART and IPTW perform poorly on SR-10.

One interesting feature of these results is that, overall, performance between the RCT and APO data is fairly similar, with similar variability ranges and most methods performing about the same.  The simulators have lower variability in general, but, for the most part, have similarly equivalent performance across methods.  This stands in contrast to the synthetic-response data sets, where we see far more variability between methods on the same data set.

The similarity between the RCT and APO data is a good sign.  APO data can be thought of as an ideal, but hard to come by, situation.  In practice, we are far more likely to only observe one potential outcome for every individual, as is the case with RCTs.  Thus, OSRCT allows us to get nearly identical results using data that is far easier to acquire than OSAPO.

This contrast between the synthetic-response data sets and the other three types has several possible explanations.  One is that the complexity of the response surface in the synthetic-response data is far higher than that of the other data sets.  Given that the response surface in the RCT data sets arise naturally in real-world systems, this suggests that the level of complexity in the synthetic-response data sets is not realistic.

Another possible explanation is that the treatment assignment in the RCT data is very simplistic (i.e., based on the value of only a single biasing covariate), while the treatment in the synthetic-response data sets is assigned based on a complex combination of many covariates.  To test this hypothesis, we defined a more complicated biasing function, using a combination of two covariates that are correlated with outcome.  Where possible, numeric covariates were chosen.  However, some data sets have only factor covariates, or a very limited number of numeric covariates, so a mix of factor and numeric biasing covariates was used.

Results with two biasing covariates are shown in Figures \ref{fig:two_bias_ATE} and \ref{fig:two_bias_rd}.  For the most part, estimates are similar to those produced with a single biasing covariate, and we still do not see the differences between algorithms that we do for the synthetic-response data sets.  It is possible that an even more complicated biasing function, potentially with many more covariates, is necessary for these results to be similar.  However, if the complexity of the treatment function is the cause of the performance difference, we would expect methods that focus on treatment modeling (such as IPTW and propensity score matching) to be performing consistently worse in this regime, which is not what we observe.


In summary, we demonstrate that three types of evaluation data (simulators, computational APO data sets, and RCTs) all produce results that are markedly different from the results produced by evaluating on synthetic-response data. Specifically, the results from these three types of evaluation data suggest that many modern methods for estimating causal effects are roughly equivalent in performance on realistic causal inference tasks.

\section{Conclusion}
Research progress in machine learning has long depended on high-quality empirical evaluation.  Research in causal inference has been hindered due to sparse empirical data resources. The growth in such data resources is slow, and the breadth of such data is still limited, especially when compared to the wealth of evaluation data sets available for associational machine learning.

Data from RCTs provides a large and growing source of data that can be used to evaluate causal inference methods.  RCT data has been widely collected by researchers in many fields over many years and is increasingly being made available for wider use.  Harnessing this wealth of data with OSRCT can substantially increase what we know about the absolute and relative performance of causal inference methods.

\section*{Acknowledgments}

Thanks to Justin Clarke for help preparing data sets and to Kaleigh Clary, Sam Witty, and Kenta Takatsu for their helpful comments. This research was sponsored by the Defense Advanced Research Projects Agency (DARPA), the Army Research Office (ARO), and the United States Air Force under under Cooperative Agreement W911NF-20-2-0005 and contracts FA8750-17-C-0l20 and HR001120C0031.  Any opinions, findings and conclusions or recommendations expressed in this document are those of the authors and do not necessarily reflect the views of DARPA, ARO, the United States Air Force, or the U.S. Government. The U.S. Government is authorized to reproduce and distribute reprints for Government purposes not withstanding any copyright notation herein.

\clearpage

\bibliography{references.bib}

\begin{thebibliography}{55}
\providecommand{\natexlab}[1]{#1}
\providecommand{\url}[1]{\texttt{#1}}
\expandafter\ifx\csname urlstyle\endcsname\relax
  \providecommand{\doi}[1]{doi: #1}\else
  \providecommand{\doi}{doi: \begingroup \urlstyle{rm}\Url}\fi

\bibitem[Arceneaux et~al.(2006)Arceneaux, Gerber, and
  Green]{arceneaux2006comparing}
Arceneaux, K., Gerber, A.~S., and Green, D.~P.
\newblock Comparing experimental and matching methods using a large-scale voter
  mobilization experiment.
\newblock \emph{Political Analysis}, pp.\  37--62, 2006.

\bibitem[Atan et~al.(2018)Atan, Jordon, and van~der Schaar]{atan2018deep}
Atan, O., Jordon, J., and van~der Schaar, M.
\newblock Deep-treat: Learning optimal personalized treatments from
  observational data using neural networks.
\newblock In \emph{Proceedings of the AAAI Conference on Artificial
  Intelligence}, volume~32, 2018.

\bibitem[Athey et~al.(2021)Athey, Imbens, Metzger, and Munro]{athey2021using}
Athey, S., Imbens, G.~W., Metzger, J., and Munro, E.
\newblock Using wasserstein generative adversarial networks for the design of
  monte carlo simulations.
\newblock \emph{Journal of Econometrics}, 2021.

\bibitem[Banzi et~al.(2019)Banzi, Canham, Kuchinke, Krleza-Jeric,
  Demotes-Mainard, and Ohmann]{Banzi2019}
Banzi, R., Canham, S., Kuchinke, W., Krleza-Jeric, K., Demotes-Mainard, J., and
  Ohmann, C.
\newblock Evaluation of repositories for sharing individual-participant data
  from clinical studies.
\newblock \emph{Trials}, 20, 2019.

\bibitem[Chipman et~al.(2007)Chipman, George, and
  McCulloch]{chipman2007bayesian}
Chipman, H.~A., George, E.~I., and McCulloch, R.~E.
\newblock Bayesian ensemble learning.
\newblock In \emph{Advances in Neural Information Processing Systems}, pp.\
  265--272, 2007.

\bibitem[Dixit et~al.(2016)Dixit, Parnas, Li, Chen, Fulco, Jerby-Arnon,
  Marjanovic, Dionne, Burks, Raychowdhury, Adamson, Norman, Lander, Weissman,
  Friedman, and Regev]{Dixit2016}
Dixit, A., Parnas, O., Li, B., Chen, J., Fulco, C.~P., Jerby-Arnon, L.,
  Marjanovic, N.~D., Dionne, D., Burks, T., Raychowdhury, R., Adamson, B.,
  Norman, T.~M., Lander, E.~S., Weissman, J.~S., Friedman, N., and Regev, A.
\newblock {Perturb-Seq: Dissecting molecular circuits with scalable single-cell
  RNA profiling of pooled genetic screens}.
\newblock \emph{Cell}, 167\penalty0 (7):\penalty0 1853--1866, 2016.

\bibitem[Dorie et~al.(2019)Dorie, Hill, Shalit, Scott, and Cervone]{Dorie2019}
Dorie, V., Hill, J., Shalit, U., Scott, M., and Cervone, D.
\newblock Automated versus do-it-yourself methods for causal inference: Lessons
  learned from a data analysis competition.
\newblock \emph{Statistical Science}, 34:\penalty0 43--68, 2019.

\bibitem[Drazen(2015)]{Drazen2015}
Drazen, J.~M.
\newblock Sharing individual patient data from clinical trials.
\newblock \emph{New England Journal of Medicine}, 372\penalty0 (3):\penalty0
  201--202, 2015.

\bibitem[Dryad(2020)]{Dryad}
Dryad, 2020.
\newblock URL \url{https://datadryad.org/stash/}.
\newblock [Online; accessed 3-June-2020].

\bibitem[Funk et~al.(2011)Funk, Westreich, Wiesen, St{\"u}rmer, Brookhart, and
  Davidian]{funk2011doubly}
Funk, M.~J., Westreich, D., Wiesen, C., St{\"u}rmer, T., Brookhart, M.~A., and
  Davidian, M.
\newblock Doubly robust estimation of causal effects.
\newblock \emph{American Journal of Epidemiology}, 173\penalty0 (7):\penalty0
  761--767, 2011.

\bibitem[Gentzel et~al.(2019)Gentzel, Garant, and Jensen]{Gentzel2019}
Gentzel, A., Garant, D., and Jensen, D.
\newblock The case for evaluating causal models using interventional measures
  and empirical data.
\newblock In \emph{Advances in Neural Information Processing Systems 32}, pp.\
  11722--11732. 2019.

\bibitem[Godlee \& Groves(2012)Godlee and Groves]{Godleee7888}
Godlee, F. and Groves, T.
\newblock The new {BMJ} policy on sharing data from drug and device trials.
\newblock \emph{British Medical Journal}, 345, 2012.

\bibitem[Guillaume \& Rougemont(2006)Guillaume and Rougemont]{Nemo}
Guillaume, F. and Rougemont, J.
\newblock Nemo: an evolutionary and population genetics programming framework.
\newblock \emph{Bioinformatics}, 22:\penalty0 2256--2557, 2006.

\bibitem[Hill(2011)]{hill2011bayesian}
Hill, J.~L.
\newblock Bayesian nonparametric modeling for causal inference.
\newblock \emph{Journal of Computational and Graphical Statistics}, 20\penalty0
  (1):\penalty0 217--240, 2011.

\bibitem[Hill et~al.(2004)Hill, Reiter, and Zanutoo]{Hill2004}
Hill, J.~L., Reiter, J.~P., and Zanutoo, E.~L.
\newblock A comparison of experimental and observational data analyses.
\newblock \emph{Applied Bayesian Modeling and Causal Inference from
  Incomplete‐Data Perspectives: An Essential Journey with Donald Rubin's
  Statistical Family}, 2004.

\bibitem[Jaciw(2016)]{jaciw2016assessing}
Jaciw, A.~P.
\newblock Assessing the accuracy of generalized inferences from comparison
  group studies using a within-study comparison approach: The methodology.
\newblock \emph{Evaluation Review}, 40\penalty0 (3):\penalty0 199--240, 2016.

\bibitem[Kallus \& Zhou(2018)Kallus and Zhou]{kallus2018confounding}
Kallus, N. and Zhou, A.
\newblock Confounding-robust policy improvement.
\newblock In \emph{Advances in Neural Information Processing Systems}, pp.\
  9269--9279, 2018.

\bibitem[Kallus et~al.(2018{\natexlab{a}})Kallus, Mao, and Udell]{Kallus2018}
Kallus, N., Mao, X., and Udell, M.
\newblock Causal inference with noisy and missing covariates via matrix
  factorization.
\newblock In \emph{Advances in Neural Information Processing Systems 31}, pp.\
  6921--6932. 2018{\natexlab{a}}.

\bibitem[Kallus et~al.(2018{\natexlab{b}})Kallus, Puli, and
  Shalit]{kallus2018removing}
Kallus, N., Puli, A.~M., and Shalit, U.
\newblock Removing hidden confounding by experimental grounding.
\newblock In \emph{Advances in Neural Information Processing Systems}, pp.\
  10888--10897, 2018{\natexlab{b}}.

\bibitem[King \& Nielsen(2019)King and Nielsen]{king2019propensity}
King, G. and Nielsen, R.~A.
\newblock Why propensity scores should not be used for matching.
\newblock \emph{Political Analysis}, 27\penalty0 (4), May 2019.

\bibitem[Kuntz et~al.(2019)Kuntz, Antman, Califf, Ingelfinger, Krumholz,
  Ommaya, Peterson, Ross, Waldstreicher, Wang, et~al.]{kuntz2019individual}
Kuntz, R.~E., Antman, E.~M., Califf, R.~M., Ingelfinger, J.~R., Krumholz,
  H.~M., Ommaya, A., Peterson, E.~D., Ross, J.~S., Waldstreicher, J., Wang,
  S.~V., et~al.
\newblock Individual patient-level data sharing for continuous learning: A
  strategy for trial data sharing.
\newblock \emph{NAM Perspectives}, 2019.

\bibitem[Kusner et~al.(2017)Kusner, Loftus, Russell, and
  Silva]{kusner2017counterfactual}
Kusner, M.~J., Loftus, J., Russell, C., and Silva, R.
\newblock Counterfactual fairness.
\newblock In \emph{Advances in Neural Information Processing Systems}, pp.\
  4066--4076, 2017.

\bibitem[LaLonde(1986)]{lalonde1986evaluating}
LaLonde, R.~J.
\newblock Evaluating the econometric evaluations of training programs with
  experimental data.
\newblock \emph{The American Economic Review}, pp.\  604--620, 1986.

\bibitem[Li et~al.(2011)Li, Chu, Langford, and Wang]{li2011unbiased}
Li, L., Chu, W., Langford, J., and Wang, X.
\newblock Unbiased offline evaluation of contextual-bandit-based news article
  recommendation algorithms.
\newblock In \emph{Proceedings of the Fourth ACM International Conference on
  Web Search and Data Mining}, pp.\  297--306, 2011.

\bibitem[Louizos et~al.(2017)Louizos, Shalit, Mooij, Sontag, Zemel, and
  Welling]{louizos2017causal}
Louizos, C., Shalit, U., Mooij, J.~M., Sontag, D., Zemel, R., and Welling, M.
\newblock Causal effect inference with deep latent-variable models.
\newblock In \emph{Advances in Neural Information Processing Systems}, pp.\
  6446--6456, 2017.

\bibitem[Mary et~al.(2014)Mary, Preux, and Nicol]{mary2014improving}
Mary, J., Preux, P., and Nicol, O.
\newblock Improving offline evaluation of contextual bandit algorithms via
  bootstrapping techniques.
\newblock In \emph{International Conference on Machine Learning}, pp.\
  172--180, 2014.

\bibitem[Miller et~al.(2020)Miller, Hsu, Troutman, Perdomo, Zrnic, Liu, Sun,
  Schmidt, and Hardt]{miller2020whynot}
Miller, J., Hsu, C., Troutman, J., Perdomo, J., Zrnic, T., Liu, L., Sun, Y.,
  Schmidt, L., and Hardt, M.
\newblock Whynot, 2020.
\newblock URL \url{https://doi.org/10.5281/zenodo.3875775}.

\bibitem[Mooij et~al.(2016)Mooij, Peters, Janzing, Zscheischler, and
  Sch{\"o}lkopf]{mooij2016distinguishing}
Mooij, J.~M., Peters, J., Janzing, D., Zscheischler, J., and Sch{\"o}lkopf, B.
\newblock Distinguishing cause from effect using observational data: Methods
  and benchmarks.
\newblock \emph{Journal of Machine Learning Research}, 17\penalty0
  (1):\penalty0 1103--1204, 2016.

\bibitem[Morgan \& Winship(2015)Morgan and Winship]{morgan2015counterfactuals}
Morgan, S.~L. and Winship, C.
\newblock \emph{Counterfactuals and causal inference}.
\newblock Cambridge University Press, 2015.

\bibitem[{NIH National Institute on Drug Abuse Data Share Website}(2020)]{NIDA}
{NIH National Institute on Drug Abuse Data Share Website}, 2020.
\newblock URL \url{https://datashare.nida.nih.gov/}.
\newblock [Online; accessed 3-June-2020].

\bibitem[Ohmann et~al.(2017)Ohmann, Banzi, Canham, Battaglia, Matei, Ariyo,
  Becnel, Bierer, Bowers, Clivio, Dias, Druml, Faure, Fenner, Galvez, Ghersi,
  Gluud, Groves, Houston, Karam, Kalra, Knowles, Krle{\v z}a-Jeri{\'c}, Kubiak,
  Kuchinke, Kush, Lukkarinen, Marques, Newbigging, O{\textquoteright}Callaghan,
  Ravaud, Schl{\"u}nder, Shanahan, Sitter, Spalding, Tudur-Smith, van Reusel,
  van Veen, Visser, Wilson, and Demotes-Mainard]{Ohmanne018647}
Ohmann, C., Banzi, R., Canham, S., Battaglia, S., Matei, M., Ariyo, C., Becnel,
  L., Bierer, B., Bowers, S., Clivio, L., Dias, M., Druml, C., Faure, H.,
  Fenner, M., Galvez, J., Ghersi, D., Gluud, C., Groves, T., Houston, P.,
  Karam, G., Kalra, D., Knowles, R.~L., Krle{\v z}a-Jeri{\'c}, K., Kubiak, C.,
  Kuchinke, W., Kush, R., Lukkarinen, A., Marques, P.~S., Newbigging, A.,
  O{\textquoteright}Callaghan, J., Ravaud, P., Schl{\"u}nder, I., Shanahan, D.,
  Sitter, H., Spalding, D., Tudur-Smith, C., van Reusel, P., van Veen, E.-B.,
  Visser, G.~R., Wilson, J., and Demotes-Mainard, J.
\newblock Sharing and reuse of individual participant data from clinical
  trials: Principles and recommendations.
\newblock \emph{BMJ Open}, 7\penalty0 (12), 2017.

\bibitem[Pearl(2019)]{pearl2019seven}
Pearl, J.
\newblock The seven tools of causal inference, with reflections on machine
  learning.
\newblock \emph{Communications of the ACM}, 62\penalty0 (3):\penalty0 54--60,
  2019.

\bibitem[{Request for Public Comment on {DRAFT} {NIH} Policy for Data
  Management and Sharing and Supplemental DRAFT Guidance}(2019)]{NIH2019}
{Request for Public Comment on {DRAFT} {NIH} Policy for Data Management and
  Sharing and Supplemental DRAFT Guidance}, 2019.

\bibitem[Rosenbaum \& Rubin(1983)Rosenbaum and Rubin]{rosenbaum1983central}
Rosenbaum, P.~R. and Rubin, D.~B.
\newblock The central role of the propensity score in observational studies for
  causal effects.
\newblock \emph{Biometrika}, 70\penalty0 (1):\penalty0 41--55, 1983.

\bibitem[Rubin(2005)]{rubin2005causal}
Rubin, D.~B.
\newblock Causal inference using potential outcomes: Design, modeling,
  decisions.
\newblock \emph{Journal of the American Statistical Association}, 100\penalty0
  (469):\penalty0 322--331, 2005.

\bibitem[Sachs et~al.(2005)Sachs, Perez, Pe'er, Lauffenburger, and
  Nolan]{sachs2005causal}
Sachs, K., Perez, O., Pe'er, D., Lauffenburger, D.~A., and Nolan, G.~P.
\newblock Causal protein-signaling networks derived from multiparameter
  single-cell data.
\newblock \emph{Science}, 308\penalty0 (5721):\penalty0 523--529, April 2005.

\bibitem[Saito \& Yasui(2020)Saito and Yasui]{saito2020counterfactual}
Saito, Y. and Yasui, S.
\newblock Counterfactual cross-validation: Stable model selection procedure for
  causal inference models.
\newblock In \emph{International Conference on Machine Learning}, pp.\
  8398--8407. PMLR, 2020.

\bibitem[Salganik et~al.(2006)Salganik, Dodds, and
  Watts]{salganik2006experimental}
Salganik, M.~J., Dodds, P.~S., and Watts, D.~J.
\newblock Experimental study of inequality and unpredictability in an
  artificial cultural market.
\newblock \emph{Science}, 311\penalty0 (5762):\penalty0 854--856, 2006.

\bibitem[Shadish et~al.(2008)Shadish, Clark, and Steiner]{Shadish2008can}
Shadish, W.~R., Clark, M.~H., and Steiner, P.~M.
\newblock Can nonrandomized experiments yield accurate answers? {A} randomized
  experiment comparing random and nonrandom assignments.
\newblock \emph{Journal of the American Statistical Association}, 103\penalty0
  (484):\penalty0 1334--1344, 2008.

\bibitem[Shalit et~al.(2017)Shalit, Johansson, and
  Sontag]{shalit2017estimating}
Shalit, U., Johansson, F.~D., and Sontag, D.
\newblock Estimating individual treatment effect: generalization bounds and
  algorithms.
\newblock In \emph{International Conference on Machine Learning}, pp.\
  3076--3085. PMLR, 2017.

\bibitem[Shi et~al.(2019)Shi, Blei, and Veitch]{shi2019adapting}
Shi, C., Blei, D.~M., and Veitch, V.
\newblock Adapting neural networks for the estimation of treatment effects.
\newblock In \emph{Advances in Neural Information Processing Systems}, pp.\
  2503--2513, 2019.

\bibitem[Shimoni et~al.(2018)Shimoni, Yanover, Karavani, and
  Goldschmnidt]{Shimoni2018}
Shimoni, Y., Yanover, C., Karavani, E., and Goldschmnidt, Y.
\newblock Benchmarking framework for performance-evaluation of causal inference
  analysis.
\newblock \emph{arXiv preprint arXiv:1802.05046}, 2018.

\bibitem[Suvarna(2015)]{Suvarna2015}
Suvarna, V.~R.
\newblock Sharing individual patient data from clinical trials.
\newblock \emph{Perspectives in Clinical Research}, 6:\penalty0 71--72, 2015.

\bibitem[Tang et~al.(2013)Tang, Rosales, Singh, and Agarwal]{tang2013automatic}
Tang, L., Rosales, R., Singh, A., and Agarwal, D.
\newblock Automatic ad format selection via contextual bandits.
\newblock In \emph{Proceedings of the 22nd ACM international conference on
  Information \& Knowledge Management}, pp.\  1587--1594, 2013.

\bibitem[Tang et~al.(2014)Tang, Jiang, Li, and Li]{tang2014ensemble}
Tang, L., Jiang, Y., Li, L., and Li, T.
\newblock Ensemble contextual bandits for personalized recommendation.
\newblock In \emph{Proceedings of the 8th ACM Conference on Recommender
  Systems}, pp.\  73--80, 2014.

\bibitem[{The Knowledge Network for Biocomplexity}(2020)]{KNB}
{The Knowledge Network for Biocomplexity}, 2020.
\newblock URL \url{https://knb.ecoinformatics.org/}.
\newblock [Online; accessed 3-June-2020].

\bibitem[Tu et~al.(2019)Tu, Zhang, Bertilson, Kjellstrom, and Zhang]{Tu2019}
Tu, R., Zhang, K., Bertilson, B., Kjellstrom, H., and Zhang, C.
\newblock Neuropathic pain diagnosis simulator for causal discovery algorithm
  evaluation.
\newblock In \emph{Advances in Neural Information Processing Systems 32}, pp.\
  12793--12804. 2019.

\bibitem[{UK Data Service}(2020)]{UKDataService}
{UK Data Service}, 2020.
\newblock URL \url{https://ukdataservice.ac.uk/}.
\newblock [Online; accessed 3-June-2020].

\bibitem[{University of Michigan Institute for Social Research}(2020)]{ICPSR}
{University of Michigan Institute for Social Research}, 2020.
\newblock URL \url{https://www.icpsr.umich.edu/web/pages/}.
\newblock [Online; accessed 3-June-2020].

\bibitem[Wager \& Athey(2017)Wager and Athey]{Wager2017}
Wager, S. and Athey, S.
\newblock Estimation and inference of heterogeneous treatment effects using
  random forests.
\newblock \emph{Journal of the American Statistical Association}, 2017.

\bibitem[Witty et~al.(2020)Witty, Takatsu, Jensen, and
  Mansinghka]{witty2020causal}
Witty, S., Takatsu, K., Jensen, D., and Mansinghka, V.
\newblock Causal inference using {G}aussian processes with structured latent
  confounders.
\newblock In \emph{International Conference on Machine Learning}, pp.\
  10313--10323. PMLR, 2020.

\bibitem[{Yale Institution for Social and Policy Studies Data
  Archive}(2020)]{ISPS}
{Yale Institution for Social and Policy Studies Data Archive}, 2020.
\newblock URL \url{https://isps.yale.edu/research/data}.
\newblock [Online; accessed 3-June-2020].

\bibitem[Yao et~al.(2018)Yao, Li, Li, Huai, Gao, and
  Zhang]{yao2018representation}
Yao, L., Li, S., Li, Y., Huai, M., Gao, J., and Zhang, A.
\newblock Representation learning for treatment effect estimation from
  observational data.
\newblock \emph{Advances in Neural Information Processing Systems}, 31, 2018.

\bibitem[Zeng et~al.(2016)Zeng, Wang, Mokhtari, and Li]{zeng2016online}
Zeng, C., Wang, Q., Mokhtari, S., and Li, T.
\newblock Online context-aware recommendation with time varying multi-armed
  bandit.
\newblock In \emph{Proceedings of the 22nd ACM SIGKDD International Conference
  on Knowledge Discovery and Data Mining}, pp.\  2025--2034, 2016.

\bibitem[Zhang \& Bareinboim(2021)Zhang and Bareinboim]{zhang2021bounding}
Zhang, J. and Bareinboim, E.
\newblock Bounding causal effects on continuous outcome.
\newblock In \emph{Proceedings of the AAAI Conference on Artificial
  Intelligence}, volume~35, pp.\  12207--12215, 2021.

\end{thebibliography}

\end{document}


\title{Supplementary Material}

\maketitle

\section{Proof of Theorem 1}
\begin{prf}
We denote the treatment for unit $i$ by $T_i$ and selected treatment based on the biasing covariates by $T^{s}_{i}$. For every unit $i$ and any treatment $t$, the biasing covariates $C^b_i$ are used to probabilistically select a treatment with probability $P(T^{s}_{i}=t|C^b_i)$.
\begin{align}
P_{D_{OSAPO}}(T_i=t) &= P(T_{i}=t)P(T^{s}_{i}=t|C^b_i)\\
&= 1*P(T^{s}_{i}=t|C^b_i)\\
\newline
P_{D_{OSRCT}}(T_i=t) &= P(T_{i}=t)P(T^{s}_{i}=t|C^b_i)\\
&= 0.5 * P(T^{s}_{i}=t|C^b_i)
\end{align}
Sub-sampling $D_{\scriptscriptstyle OSAPO}$ uniformly at random is equivalent to multiplying $P_{D_{OSAPO}}(T_i=t)$ by a scaling factor, $s$.  When $s = 0.5$, $P_{D_{OSRCT}}(T_i=t) = P_{D_{OSAPO}}(T_i=t)$.
\end{prf}

\section{Proof of Theorem 2}

\begin{proof}
Assume binary treatment $T \in \{0,1\}$.  For any unit $i$ with biasing covariates $C^b_i$, let $P(T_i=t)=p_t$, $P(T^{s}_{i}=t|C^b_i)=p_{T^{s}_{i}=t|c^b}$, and $n=|D_{\scriptscriptstyle RCT}|$.  Indices are omitted when clear from context.
\begin{align*}
    P(i \in D_{\scriptscriptstyle OSRCT}) &= p_1 p_{T^{s}_{i}=1|c^b} + p_0 p_{T^{s}_{i}=0|c^b}\\
    &= p_1 p_{T^{s}_{i}=1|c^b} + (1-p_1) (1-p_{T^{s}_{i}=1|c^b})\\
    &= 2p_1 p_{T^{s}_{i}=1|c^b} - p_1 - p_{T^{s}_{i}=1|c^b} + 1\\
    E[|D_{\scriptscriptstyle OSRCT}|] &= \sum_{i=1}^{n}[2p_1 p_{T^{s}_{i}=1|c^b} - p_1 - p_{T^{s}_{i}=1|c^b} + 1]\\
    &= n-np_1+(2p_1-1)\sum_{i=1}^{n}p_{T^{s}_{i}=1|c^b}
\end{align*} 
If either $p_1 = 0.5$ or $\sum_{i=1}^{n}p_{T^{s}_{i}=1|c^b} = 0.5n$, $E[|D_{\scriptscriptstyle OSRCT}|] = 0.5n$.
\end{proof}

\section{Using the Complementary Sample for Evaluation}
\label{section:complementary_sample}
One challenge when evaluating causal inference methods on their ability to estimate unit-level effects of interventions is the need for a held-out test set. The constructed observational data is constructed by sub-sampling the original RCT data.  This means that evaluating on all of the RCT data may produce a biased result by testing on a superset of the training data.  One potential solution is to divide the RCT data into separate training and test sets.  However, since OSRCT necessarily reduces the size of the training data by sub-sampling, the extra requirement of holding out a test set limits the number of RCTs that can be used, since not all randomized experiments will have enough data to learn effective models after two rounds of sub-sampling.

A more data-efficient approach is to use the data rejected by the biased sub-sampling.  OSRCT sub-samples RCT data to create a probabilistic dependence between the biasing covariates and treatment.  Based on the values of the biasing covariates, a treatment is selected for every unit.  If that treatment is present in the data, the unit is included in the sample; otherwise the unit is rejected.  This rejected sample (which we call the \textit{complementary sample}) also has a causal dependence from the biasing covariates to treatment.  The only difference is that the form of that dependence is the complement of that for the accepted sample, such that covariate values that lead to a high probability of treatment in the accepted sample would lead to a low probability of treatment in the complementary sample.  Because we know the functional form of this induced bias, we can weight the data points in the complementary sample  according to their probability of being included in the accepted sample.  In aggregate, this type of weighting allows the complementary sample to approximate the distribution of the training data, and thus be used for testing.  This is equivalent to inverse propensity score weighting \citep{rosenbaum1983central}.

\begin{manualtheorem}{3}
\label{thm:complementary-sample}
For binary treatment $T \in \{0,1\}$, biasing covariates $C^{b}$, outcome $Y$, estimated outcome $\hat{Y}$, biased sample \DOSRCT\ and complementary sample $\bar{D}_{\scriptscriptstyle OSRCT}$, let $p_s = P(T^{s}_{i}=t_i|C^{b}_{i})$.  Then, $E[\hat{Y}-Y]$ for \DOSRCT\ = $E[(\hat{Y}-Y)\frac{p_s}{1-p_s}]$ for $\bar{D}_{\scriptscriptstyle OSRCT}$.
\end{manualtheorem} 

\begin{proof}
For \DOSRCT, 
\begin{align*}
    E[\hat{Y}-Y]_{D_{\scriptscriptstyle OSRCT}} = E[P(T^{s}_{i}=t_i|C^{b}_{i})(\hat{Y_i}-Y_i)]\\
\end{align*}
For $\bar{D}_{\scriptscriptstyle OSRCT}$,
\begin{align*}
    E[\hat{Y}-Y]_{\bar{D}_{\scriptscriptstyle OSRCT}} = E[(1-P(T^{s}_{i}=t_i|C^{b}_{i}))(\hat{Y_i}-Y_i)]  
\end{align*}
If we weight the outcome estimates for $\bar{D}_{\scriptscriptstyle OSRCT}$ by $\frac{P(T^{s}_{i}=t_i|C^{b}_{i})}{1-P(T^{s}_{i}=t_i|C^{b}_{i})}$,
\begin{align*}
    E[\hat{Y}-Y]_{\bar{D}_{\scriptscriptstyle OSRCT}} &= E[\frac{P(T^{s}_{i}=t|C^{b}_{i})}{1-P(T^{s}_{i}=t_i|C^{b}_{i})}\cdot\\
    (1&-P(T^{s}_{i}=t_i|C^{b}_{i}))(\hat{Y_i}-Y_i)]\\
    &= E[P(T^{s}_{i}=t_i|C^{b}_{i})(\hat{Y_i}-Y_i)]\\
    &= E[\hat{Y}-Y]_{D_{\scriptscriptstyle OSRCT}}
\end{align*}
\end{proof}

\section{Experimental Evaluation of Theorems 1 and 3}

Intuitively, the procedure outlined in Algorithm 2 works because treatment is \textit{randomly} assigned in RCTs.  The data is sub-sampled based solely on the value of a probabilistic function of the biasing covariates, which selects a value of treatment for every unit $i$.  Since the observed treatment is randomly assigned, it contains no information about any of $i$'s covariates.  The only bias introduced by this sub-sampling procedure is the intended bias: a particular form of causal dependence from $C^{b}$ to $T$.

To assess OSRCT's effectiveness at approximating APO data, we performed an experiment using an APO data set provided by \citet{Gentzel2019}, replicating their experimental setup.  In this data, units are Postgres queries, interventions are Postgres settings (such as type of indexing), covariates are features of queries (such as the number of joins or the number of rows returned), and outcomes are measured results of running the query (such as runtime).  If the Postgres database is queried in a recoverable manner, the same query can be run repeatedly while varying the treatment, creating APO data.  For this analysis, consistent with \citet{Gentzel2019}, we chose runtime as the outcome, indexing level as the treatment, and the number of rows returned by the query as the biasing covariate.

\begin{figure*}[h]
\centering
\includegraphics[width=.87\textwidth]{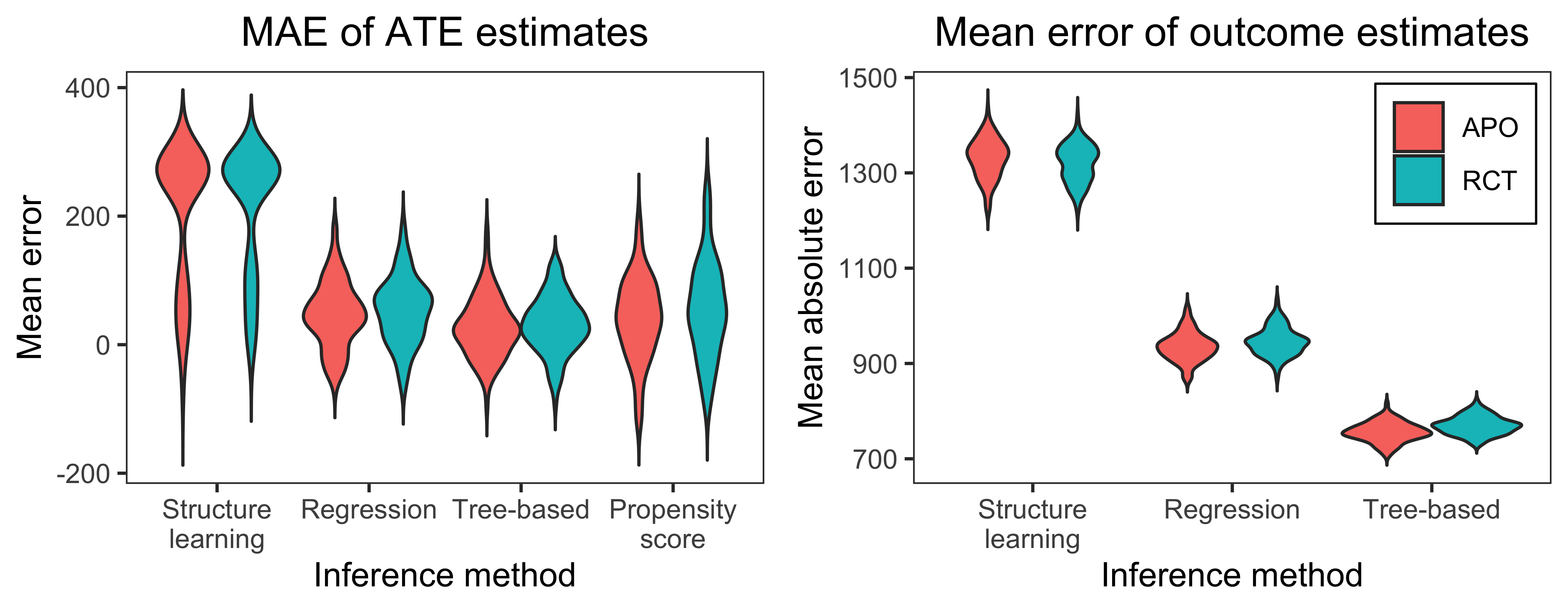}
\caption{APO vs RCT sampling on Postgres data. \textit{Left}: Mean absolute error of ATE estimates, \textit{Right}: Mean error of estimated outcomes The similarity between the RCT and APO data sets suggests that OSRCT and OSAPO produce equivalent constructed observational data.}
\label{fig:postgres-results}
\end{figure*}

To compare RCT and APO data, We converted the APO Postgres data into RCT-style data by randomly sampling a single treatment for every unit.  We then created constructed observational data from both the original APO data and the RCT-style data, creating \DOSAPO\ and \DOSRCT.  For \DOSRCT, as described in Theorem 3, outcome estimation was evaluated by weighting the errors in the complementary sample.  However, in \DOSAPO, no complementary sample is created, since the selected treatment is guaranteed to be observed for every unit.  Instead, we can divide \DOSAPO\ into training and test sets.  If the RCT-style data is created by sub-sampling treatments equally, by Theorem 2, splitting \DOSAPO\ in half leads to a data set approximately the same size as \DOSRCT, allowing for comparison with equal training set size.  We estimated errors over 100 trials.  Results are shown in Figure \ref{fig:postgres-results}.

Results are very similar for the APO data and the RCT-style data constructed from it.  Consistent with Theorem 1, this suggests that evaluation with OSRCT data produces equivalent results to OSAPO data.  In addition, consistent with Theorem 3, the similarity in outcome estimates suggests that weighting the complementary sample produces equivalent results to an unweighted held-out test set.

\begin{table*}[t]
\hspace*{-.7cm}
\caption[Data sets used in experiments.]{\normalsize Data sets used in experiments. `ID' denotes the repository-specific ID for each data set, where applicable.  `Coding' denotes the shortened data set name used in figures.}
\centering
\resizebox{1.1\textwidth}{!}{%
\begin{tabular}{c | c | c | c | c | c | c | c}
\textbf{Source} & \textbf{ID} & \textbf{Coding} & \textbf{Sample} & \textbf{Num} & \textbf{Treatment} & \textbf{Outcome} & \textbf{Biasing}\\
 & & & \textbf{Size} & \textbf{Covars} & & & \textbf{Covar}\\
\hline
Dryad & 4f4qrfj95 & RCT-1 & 6453 & 27 & Temperature & Plant health & Species \\
Dryad & B8KG77 & RCT-2 & 15289 & 4 & Video type & Bicycle rating & Bike access \\
HDV & WT4I9N & RCT-3 & 551 & 5 & Fact truth & Fact removed & Fact cited \\
ICPSR & 20160213 & RCT-4 & 5573 & 10 & Guest race & Accepted & Prior black tenants \\
ICPSR & 23980 & RCT-5 & 10098 & 7 & Age & Resume response & Volunteer service \\
ISPS & d037 & RCT-6 & 4859 & 2 & Race & Legislator response & Party \\
ISPS & d084 & RCT-7 & 48509 & 6 & E-mail source & Voter turnout & Prior election turnout \\
ISPS & d113 & RCT-8 & 10200 & 4 & Mailing & Voter turnout & Gender \\
KNB & 1596312 & RCT-9 & 760 & 4 & Soil heating & C02 levels & Depth \\
KNB & f1qf8r51t & RCT-10 & 8063 & 4 & Plant protection & Plant survival & Location \\
musiclab & - & RCT-11 & 3719 & 13 & peer-influence & average rating & music knowledge \\
NIDA & P1S1 & RCT-12 & 776 & 5 & Nicotine levels & Cigarettes per day & Weight \\
UK Data Service & 852874 & RCT-13 & 343 & 5 & Shown video & Response & Ethnicity \\
UK Data Service & 853369 & RCT-14 & 4210 & 3 & Biasing instruction & Line-up identification & Recruitment method \\
UK Data Service & 854092 & RCT-15 & 691 & 5 & Fact check validity & Reaction & Political activity \\
JDK & - & APO-1 & 473 & 5 & Obfuscate & Num bytecode ops & Test javadocs \\
Networking & - & APO-2 & 2599 & 1 & Proxy & Elapsed time & Server class \\
Postgres & - & APO-3 & 11128 & 8 & Index level & Runtime & Rows returned \\
Nemo & - & Sim-1 & 10000 & 9 & Breeding & Adult viability & Deleterious loci\\
Nemo & - & Sim-2 & 10000 & 9 & Deleterious model & Deleterious frequency & Mutation rate \\
Nemo & - & Sim-3 & 10000 & 10 & Dispersal rate & Survival & Deleterious loci \\
Neuropathic pain & - & Sim-4 & 10000 & 25 & DLS L4-L5 & Lumbago & DLS L5-S1 \\
Neuropathic pain & - & Sim-5 & 10000 & 25 & DLS C5-C6 & Right Skull pain & DLS C3-C4 \\
Neuropathic pain & - & Sim-6 & 10000 & 25 & DLS C4-C5 & Right Shoulder pain & DLS C6-C7 \\
Whynot & opiod & Sim-7 & 10000 & 3 & Abuse & Overdose deaths & Illicit users \\
Whynot & world2 & Sim-8 & 10000 & 6 & Capital investiment & Population & Pollution \\
Whynot & zika & Sim-9 & 10000 & 9 & Zika control strategy & Symptomatic humans & Exposed mosquitoes \\
\end{tabular}}
\hspace{1cm}
\label{tbl:datasets}
\end{table*}

\begin{table*}[t]
\hspace*{-.7cm}
\caption[ACIC data sets used in experiments.]{ACIC Data sets used in experiments. `ID' denotes the ACIC ID for each data set.  `Coding' denotes the shortened data set name used in figures.}
\centering
\resizebox{1\textwidth}{!}{%
\begin{tabular}{c | c | c | c | c | c | c | c | c}
\textbf{ID} & \textbf{Coding} & \textbf{Sample} & \textbf{Num} & \textbf{Treatment} & \textbf{Percent} & \textbf{Outcome} & \textbf{Alignment} & \textbf{Treatment Effect}\\
 & & \textbf{Size} & \textbf{Covars} & \textbf{Function} & \textbf{Treated} & \textbf{Function} & & \textbf{Heterogeneity} \\
\hline
4 & SR-1 & 4802 & 56 & Polynomial & 35\% & Exponential & 75\% & high \\
27 & SR-2 & 4802 & 56 & Polynomial & 35\% & Step & 25\% & Medium \\
47 & SR-3 & 4802 & 56 & Polynomial & 65\% & Exponential & 75\% & High \\
65 & SR-4 & 4802 & 56 & Step & 65\% & Step & 75\% & Medium \\
71 & SR-5 & 4802 & 56 & Step & 65\% & Step & 25\% & High \\
\end{tabular}}
\hspace{1cm}
\label{tbl:acic-datasets}
\end{table*}

\begin{table*}[t]
\hspace*{-.7cm}
\caption[IBM data sets used in experiments.]{IBM Data sets used in experiments. `ID' denotes the IBM ID for each data set.  `Coding' denotes the shortened data set name used in figures.}
\centering
\resizebox{1\textwidth}{!}{%
\begin{tabular}{c | c | c | c | c | c | c}
\textbf{ID} & \textbf{Coding} & \textbf{Sample} & \textbf{Num} & \textbf{Percent} & \textbf{Effect} & \textbf{Link} \\
 & & \textbf{Size} & \textbf{Covars} & \textbf{Treated} & \textbf{Size} & \textbf{Type} \\
\hline
1b50aae9f0e34b03bdf03ac195a5e7e9 & SR-6 & 10000 & 151 & 69\% & -3.2 & Polynomial \\
2b6d1d419de94f049d98c755beea4ae2 & SR-7 & 10000 & 151 & 23\% & -0.13 & Log \\
19e667b985624159bae940919078d55f & SR-8 & 10000 & 151 & 17\% & 0.06 & Exponential \\
7510d73712fe40588acdb129ea58339b & SR-9 & 10000 & 151 & 27\% & 0.017 & Log \\
c55cbee849534815ba80980975c4340b & SR-10 & 10000 & 151 & 19\% & -0.23 & Exponential \\
\end{tabular}}
\hspace{1cm}
\label{tbl:ibm-datasets}
\end{table*}

\section{Details about RCT Repositories}

We selected data sets from six repositories: (1) Dryad \citep{Dryad}; (2) the Yale Institution for Social and Policy Studies Repository \citep{ISPS}; (3) the NIH National Institute on Drug Abuse Data Share Website \citep{NIDA}; (4) the University of Michigan's ICPSR repository \citep{ICPSR}; (5) the UK Data Service \citep{UKDataService}; and (6) the Knowledge Network for Biocomplexity \citep{KNB}.  These repositories were selected because they contained RCT data, were reasonably well-documented, and had a simple access process.  None of these repositories house RCT data exclusively, so some search and filtering was necessary to identify relevant data sets.

Many other repositories exist that contain RCT data but have higher access restrictions.  Access to these repositories generally involves requesting permission for any desired data set.  For some, this request only involves submitting a brief description of the intended use and proving sufficient credentials.  For others, this request may require a detailed data analysis plan and description of the benefits of the research.  Examples include the National Institute of Diabetes and Digestive and Kidney Diseases \citep{NIDDK}, Vivli \citep{Vivli}, The National Institute of Mental Health Data Archive \citep{NDA}, Project Data Sphere \citep{ProjectDataSphere}, and the Data Observation Network for Earth \citep{DataOne}.

\section{Details about RCT Data Sets Used in Demonstration}

The data sets used in the demonstration came from six repositories, all of which allowed for direct download of the data after registering with the repository.  Each of these data sets met five criteria:
\begin{description}
    \item[Random assignment:] Treatment must be fully randomized for OSRCT to work as intended.  We ensured that the selected data sets were created by randomly assigning treatment to each unit.
    \item[Independent units:] Many causal inference methods assume independent data instances, so we ensured that the units in the data sets could reasonably be assumed independent (e.g., no spatial correlation).
    \item[Measured pre-treatment covariate:] At least one measured pre-treatment covariate is necessary to induce confounding bias.  The data sets we selected all had multiple pre-treatment covariates, allowing us to select one that was correlated with outcome to induce confounding bias.
    \item[Reasonably large sample size:] Many RCT data sets are very small ($N<100$).  We selected only reasonably large data sets ($N>500$).
    \item[Ease of use:] Some data sets were poorly documented or stored the data over many files.  We selected data sets that would require minimal pre-processing.
\end{description}

In cases where treatment was not binary, a reasonable binary version of treatment was constructed, either by grouping merging treatment categories or by selecting a subset of the data with only two values of treatment.  Details about these data sets are given in Table \ref{tbl:datasets}.

We also used some additional data set for the evaluation: synthetic-response data sets from the ACIC Competition and the IBM Causal Inference Benchmarking Framework \cite{Dorie2019, Shimoni2018}, APO data sets from computational systems \cite{Gentzel2019}, and three simulators \cite{Nemo, Tu2019, miller2020whynot}.  Details about these data sets can be found in Tables \ref{tbl:datasets}, \ref{tbl:acic-datasets}, and \ref{tbl:ibm-datasets}.  For the experimental results presented in the paper, 5000 samples were used for the IBM Causal Inference Benchmarking Framework data sets, rather than 2000, due to the high number of covariates.

\section{Details about Causal Inference Methods Evaluated}

For each causal inference method evaluated, we used the default implementation.  While it may be possible to achieve better performance for some of the these methods after parameter tuning, we focused our evaluation on default performance rather than best-case.  Many practitioners looking to apply a causal inference method will default to using the initial parameter settings of a method, so it is a useful case to compare.  Our comparison also includes 37 data sets, and individually tuning parameters for seven algorithms across 37 data sets was beyond the scope of this paper.  A deep dive into the performance of any individual causal inference method, varying parameter settings and implementation details, is a possible avenue for future work and could produce interesting results.

\textbf{Propensity-score matching (PSM)} learns a model of treatment probability that is used to produce samples with equal probabilities of treatment. Then weighted outcome estimates of the treatment and control populations are compared to estimate ATE.  PSM was implemented using the \textit{MatchIt} package in R.

\textbf{Inverse probability of treatment weighting} (IPTW) is similar to propensity score matching, in that both estimate the probability of treatment and use that to control for confounding.  Rather than using the probability of treatment to match individuals between the treatment and control populations, IPTW \textit{weights} the outcomes of every individual according to their probability of treatment and uses these weighted outcomes to estimate ATE \cite{rosenbaum1983central}.

\textbf{Outcome regression} is one simple approach for effect estimation that models outcome given treatment and all measured covariates.  Unlike the potential outcomes approaches discussed above, outcome regression makes no attempt to model the treatment mechanism, focusing solely on effectively modeling outcome.  Recent studies have suggested that effectively modeling outcome may be more important than trying to account for differences in treatment assignment \cite{Dorie2019}.

\textbf{Bayesian Additive Regression Trees} (BART) use a tree-based model to estimate the response surface, allowing for estimation of both ATE and individual outcomes \cite{chipman2007bayesian}. Regression trees are a type of tree used when the outcome is continuous, which partition the input data into subgroups with similar outcomes.  BART creates an ensemble of sequentially-learned regression trees, with a regularization prior to keep the effects of individual trees small.  Estimates for the ensemble are obtained by summing the outputs of all the trees.  When used for causal modeling, all observed covariates and treatment are used as predictors of outcome, and estimates of ATE can be obtained by estimating outcome for all individuals with both $T=1$ and $T=0$ and calculating the mean difference.  BART was implemented using the \textit{bartMachine} R package.

\textbf{Causal forests} are random forests that specifically estimate ATE \cite{Wager2017}. They make use of causal trees \cite{athey2016recursive}, which estimate ATE at the leaf nodes by splitting such that the the number of training points at the leaf node is small enough to be treated as though they came from a randomized experiment.  A causal forest then averages the ATE estimates from the causal trees in the ensemble to get an overall estimate of ATE.  This was implemented using the \textit{grf} R package, with the default parameters.

The above methods focus on modeling either treatment or outcome.  However, if this model is misspecified, the effect estimate will be biased.  \textbf{Doubly-robust methods} are designed to avoid this issue, producing an unbiased estimate of ATE as long as either the treatment or the outcome model is correctly specified.  This is commonly implemented as a combination of IPTW weighting and outcome regression \cite{funk2011doubly}.  Doubly-robust estimation was implemented using the \textit{fastDR} R package.

Shi et al.~\cite{shi2019adapting} propose a neural-network-based method, using a new proposed architecture called \textbf{Dragonnet}.  This approach uses a deep neural network to produce a representation layer of the covariates.  This representation layer is then used to predict both treatment and outcome.  The prediction of treatment acts as a propensity score, which is used to adjust for confounding when estimating treatment effect.  Dragonnet net is one example of a class of neural-network-based approaches for causal modeling, which generally follow a similar approach. \cite{johansson2016learning, shalit2017estimating, schwab2018perfect, louizos2017causal, yoon2018ganite}

\section{Additional results}

\subsection{Neural-network results}
\begin{figure*}
\centering
\includegraphics[width=.7\textwidth]{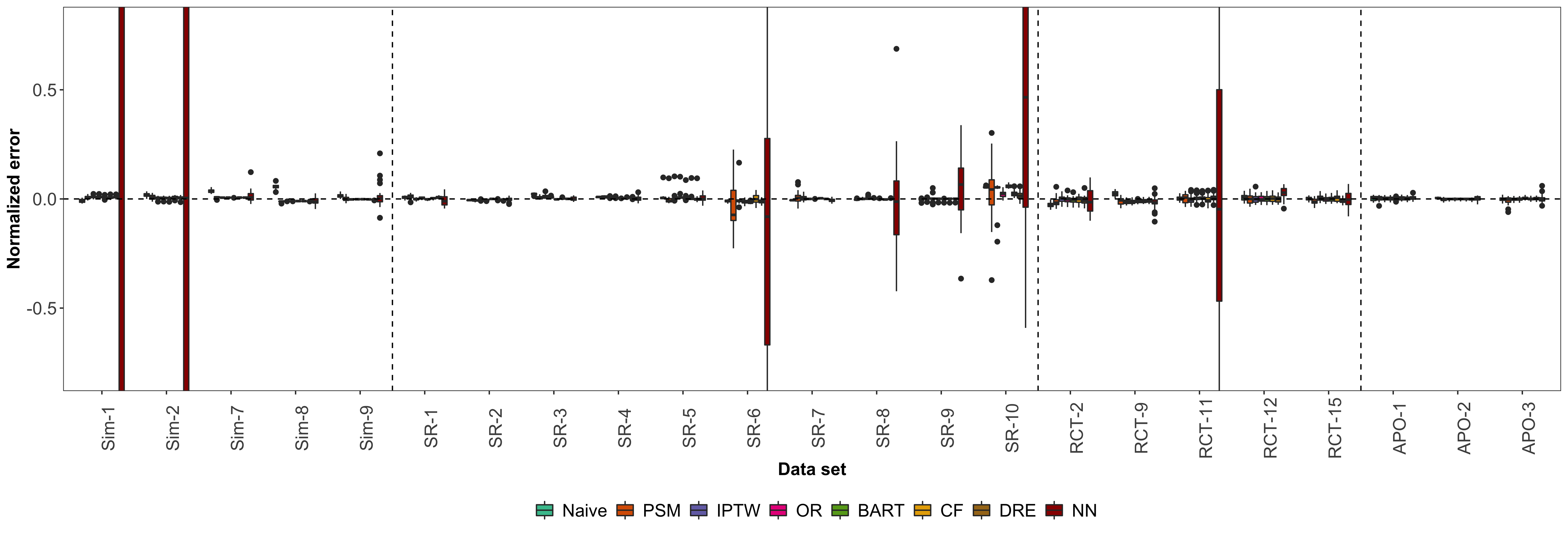}
\caption{Normalized error in estimating ATE for data sets with continuous outcome}
\label{fig:nn_ATE}
\end{figure*}

\begin{figure*}
\centering
\includegraphics[width=.7\textwidth]{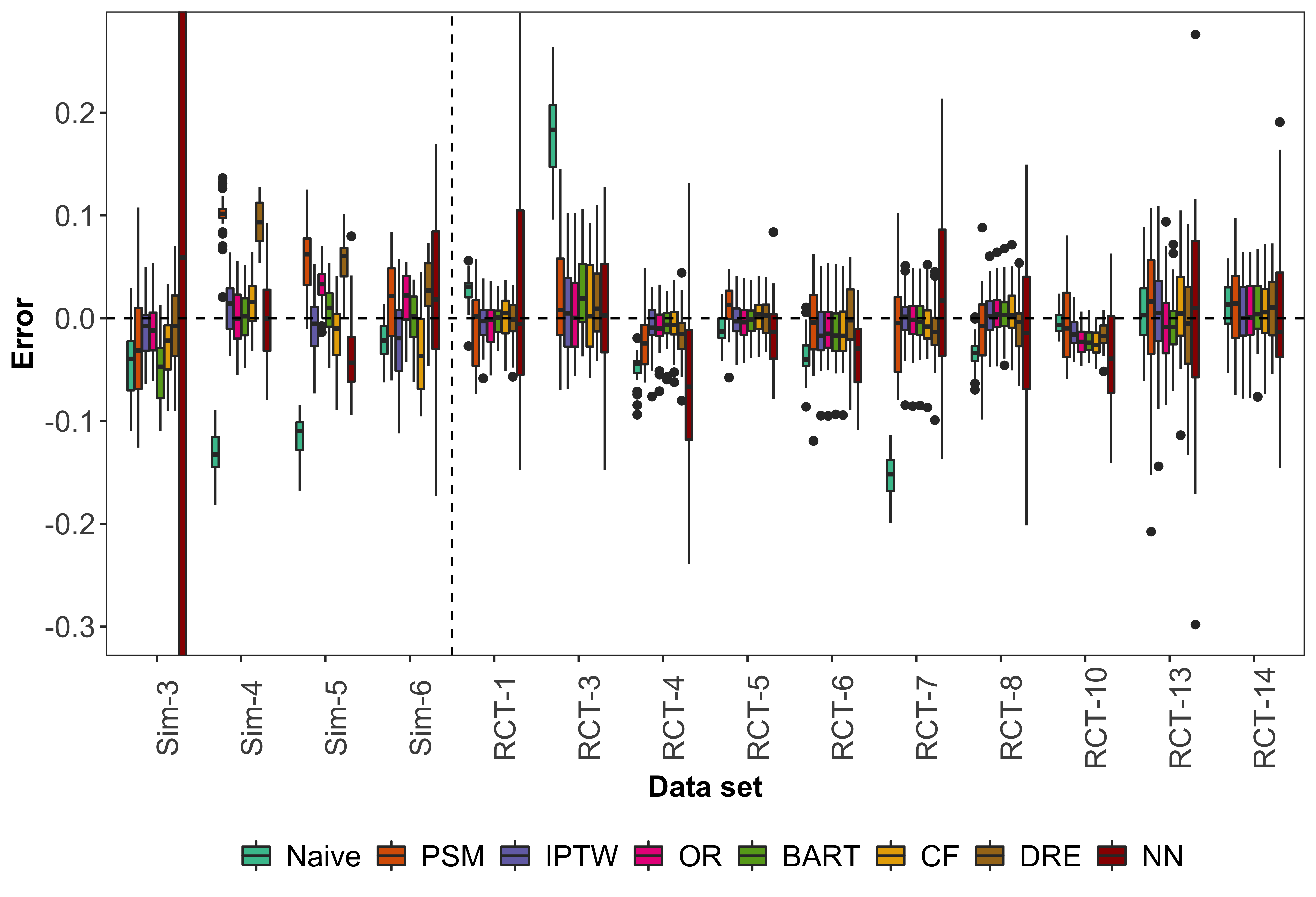}
\caption{Error in estimating risk difference for data sets with binary outcome}
\label{fig:nn_rd}
\end{figure*}

We also ran experiments comparing against a neural-network-based method \cite{shi2019adapting}. \footnote{We tried neural-network method implementations by Shi et al.: Dragonnet, Dragonet + TMLE, Tarnet, and Tarnet+TMLE.  The performance was comparable between all four, so only results for Dragonnet+TMLE are reported.} This method has significantly higher variability than the other methods.  There are a couple of possible explanations for this.  As we initialize different random weights in each run, the model might be sensitive to the initialization weights and converge to different local optima. In addition, sample size for most of the data sets is less than 5000, which is significantly lower than is typically used for neural network based methods. This might be causing overfitting and high variability.  We ran the neural network method using the experimental set up described in the main paper, (as in Figures 3 and 4 in the main paper).  The results are shown in Figures \ref{fig:nn_ATE} and \ref{fig:nn_rd}.

\subsection{Outcome estimation results}
\begin{figure}
\centering
\includegraphics[width=.45\textwidth]{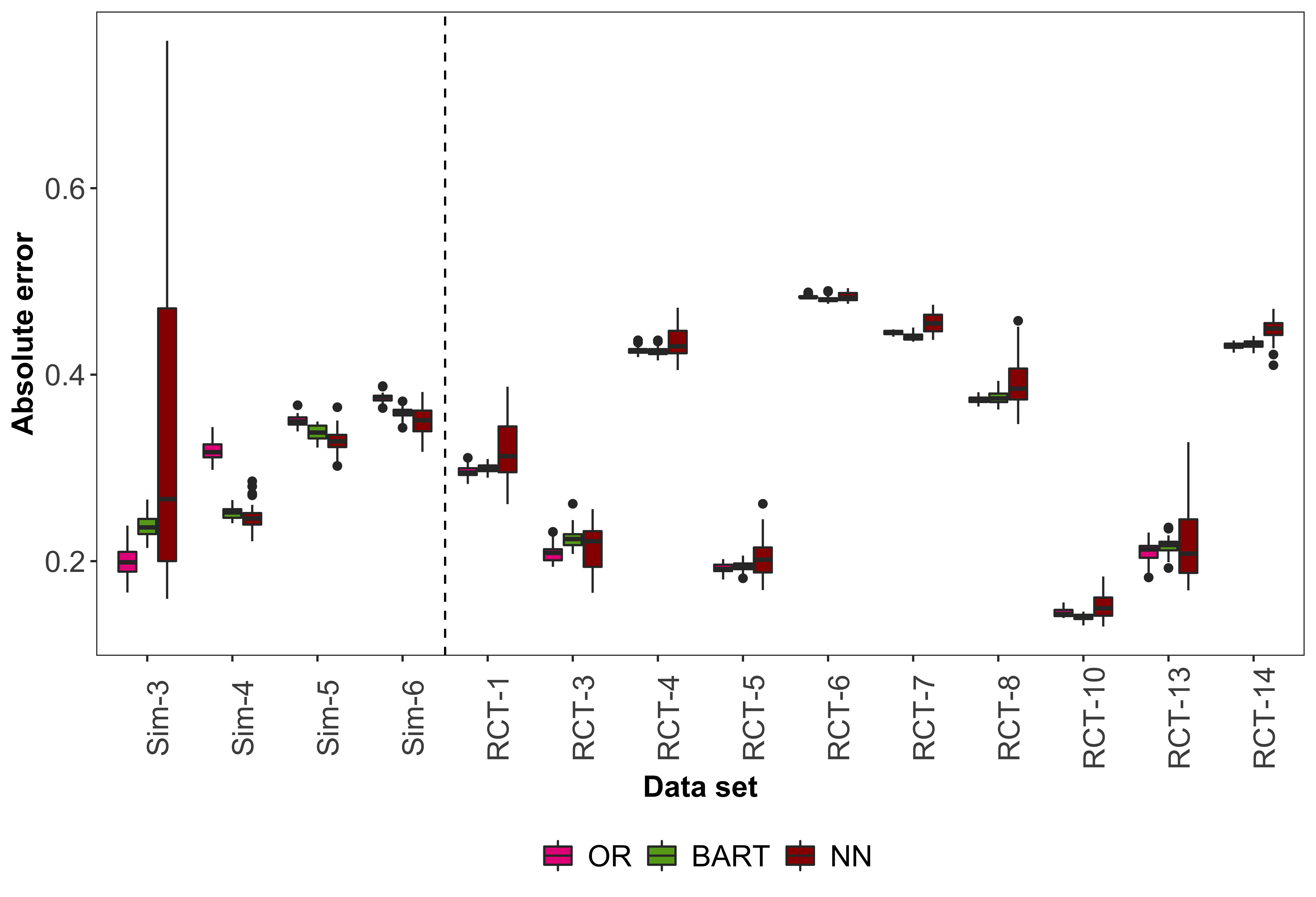}
\caption{Error in estimating a binary outcome}
\label{fig:initial_outcome_binary}
\end{figure}

\begin{figure*}
\centering
\includegraphics[width=.9\textwidth]{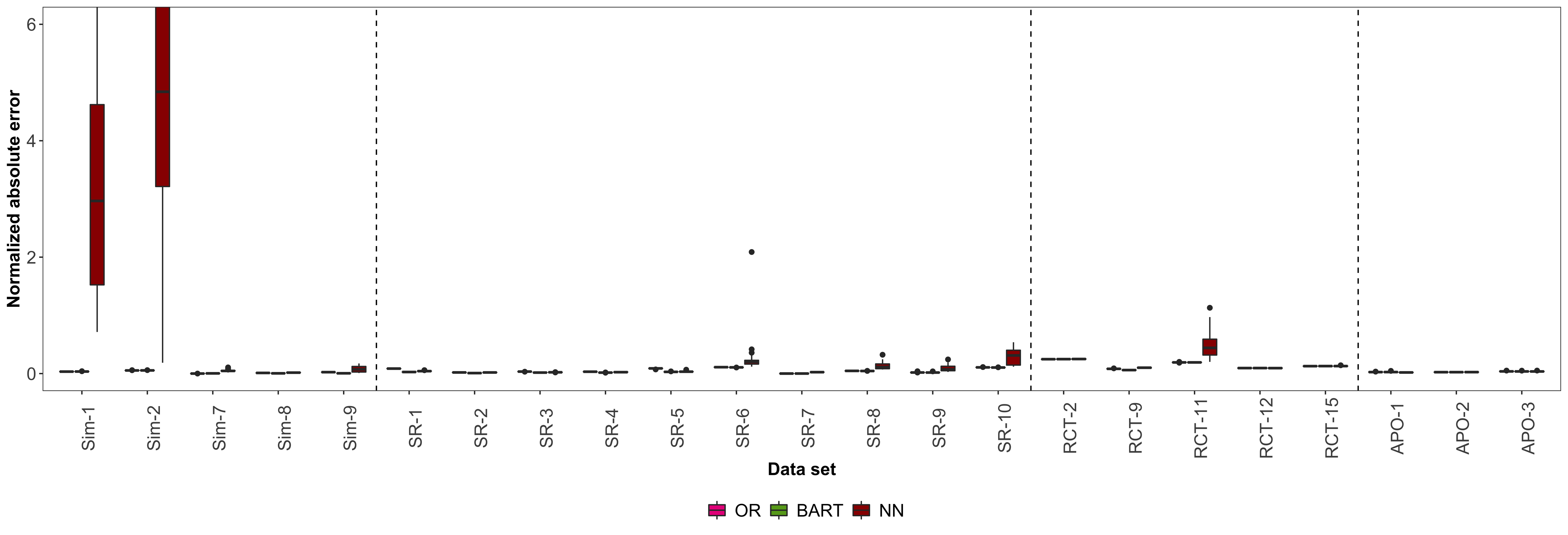}
\caption{Normalized absolute error in estimating a continuous outcome}
\label{fig:initial_outcome_cont}
\end{figure*}

\begin{figure*}
\centering
\includegraphics[width=.9\textwidth]{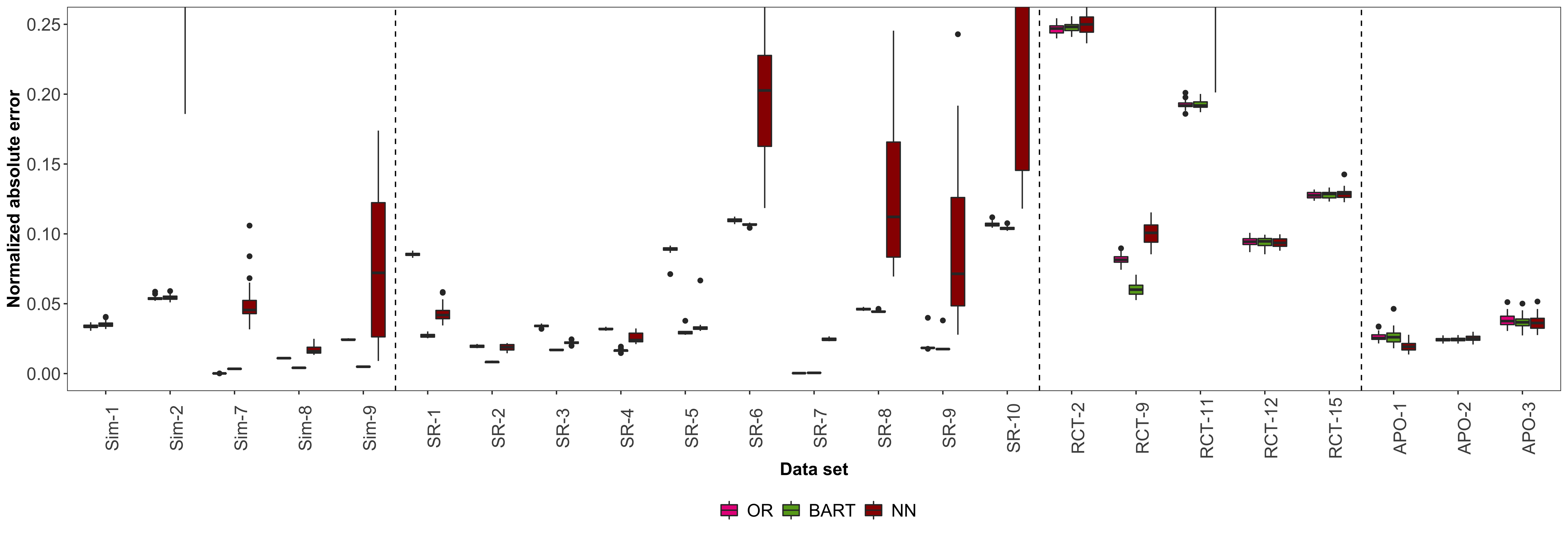}
\caption{Normalized absolute error in estimating a continuous outcome, zoomed in to show details}
\label{fig:initial_outcome_cont_zoomed}
\end{figure*}

For APO data sets and synthetic response data sets, a held out test set can be used instead, which, by Theorem \ref{thm:complementary-sample}, is equivalent to using the weighted complementary sample.  However, many of the algorithms we are evaluating here are not capable of producing individual-level outcome estimates, so this evaluation is limited to only BART,  outcome regression., and the neural-network method.  Results for data sets with binary outcomes are shown in Figure \ref{fig:initial_outcome_binary}, and results for data sets with continuous outcomes are shown in Figure \ref{fig:initial_outcome_cont}. (Figure \ref{fig:initial_outcome_cont_zoomed} contains the continuous outcome results but zoomed in, cutting off some extreme neural network results to show details).

Unsurprisingly, BART consistently outperforms outcome regression.  Both of these methods focus on modeling the response surface, but BART uses a higher capacity tree-based model rather than a simple regression.  The difference is far stronger for data sets with a continuous outcome, compared to those with a binary outcome.  The difference is also minimal for the RCT data sets.  This trend is constant as we increase the strength of the biasing and when two biasing covariates are used.  The neural network has significantly higher variability than BART and outcome regression.  Since no hyperparameter tuning was performed for the neural-network method, and many data sets have low sample size, it is not surprising that the neural-network method sometimes does very poorly.  However, in many cases, the mean error is similar, and for many data sets with continuous outcome, performance is equivalent to BART and outcome regression. This evaluation is unfortunately limited since none of the other algorithms we evaluated are capable of producing individual-level outcome estimates.  In general, methods that model outcome are more likely to provide this as an option, making this a useful evaluation tool when comparing multiple outcome estimation-based methods.

\subsection{Comparison between Sub-sampling and Weighting}
\begin{figure*}
\centering
\includegraphics[width=.9\textwidth]{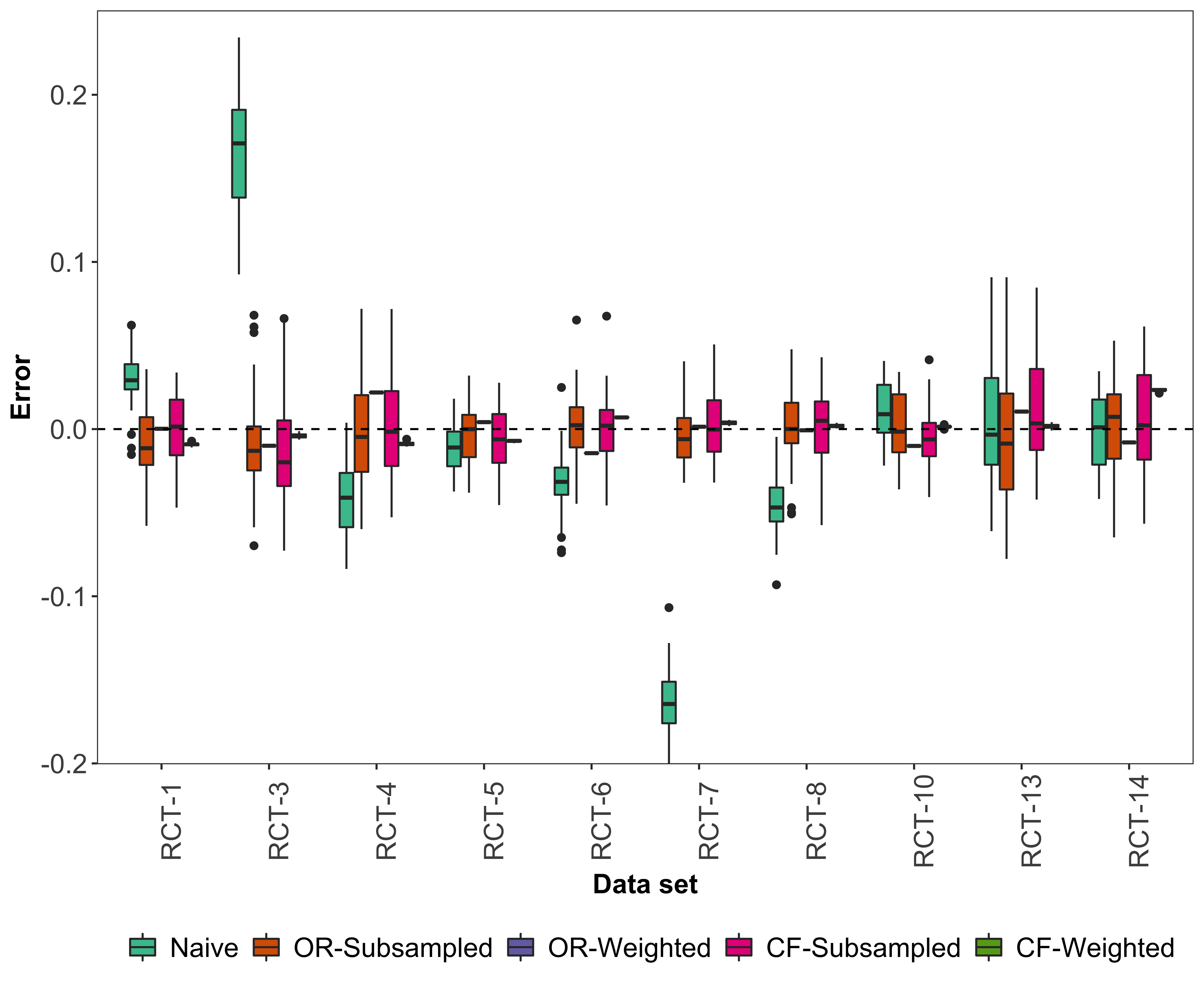}
\caption{Comparison between weighting and sub-sampling approach in estimating a binary outcome}
\label{fig:sampling_weighting_risk_ratio}
\end{figure*}

\begin{figure*}
\centering
\includegraphics[width=.9\textwidth]{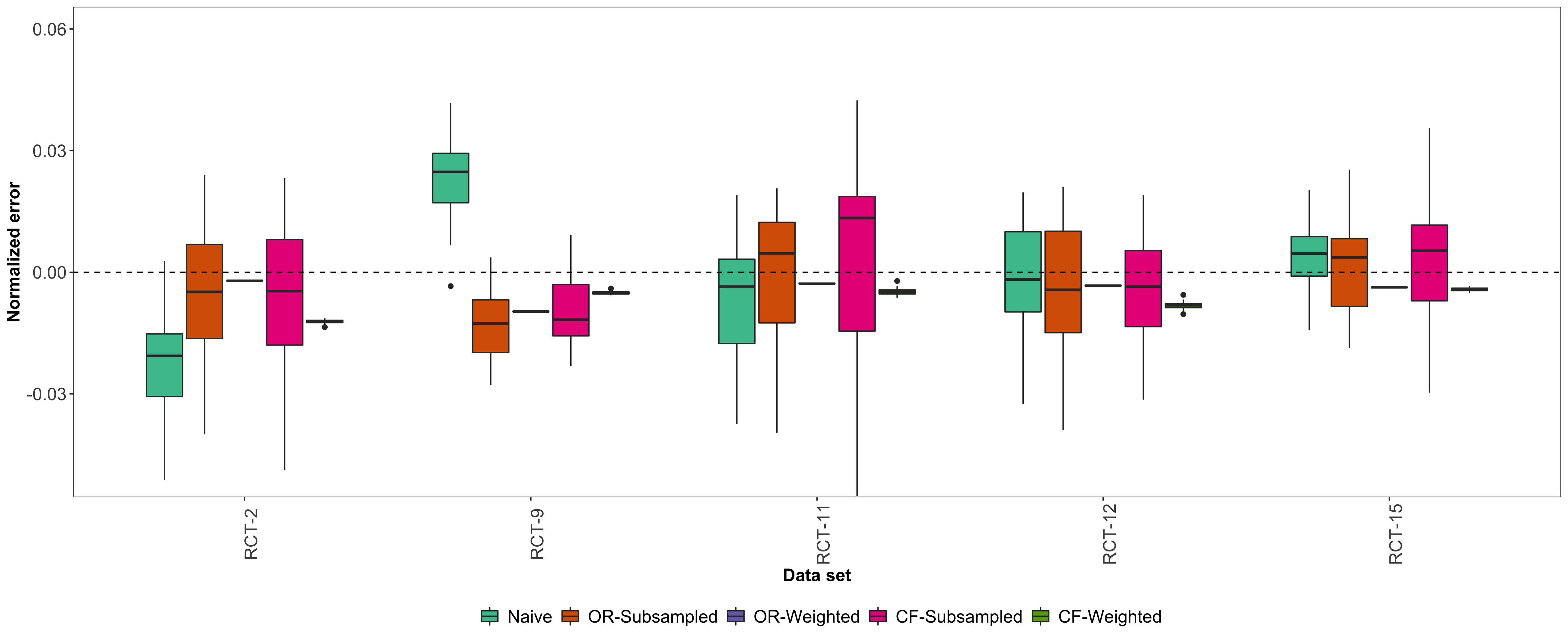}
\caption{Comparison between weighting and sub-sampling approach in estimating a continuous outcome}
\label{fig:sampling_weighting_ATE}
\end{figure*}

An alternative approach to sub-sampling, as mentioned in Section 2.1 of the paper, is to reweight the units, according to $P(T^{s}_{i}=t_i|C^{b}_{i})$.  This approach requires that the causal inference method under evaluation accepts unit-level weights. Among the methods we used for evaluation, only \textit{causal forests} and \textit{outcome regression} accept unit-level weights. To assess the similarity between the two approaches, we compared the causal effect estimates obtained using sub-sampling to those obtained using weighting. Results for binary outcomes can be seen in Figure \ref{fig:sampling_weighting_risk_ratio}, and results for continuous outcomes can be seen in Figure \ref{fig:sampling_weighting_ATE}. Naive, OR-subsampled, CF-subsampled estimates are calculated using the sub-sampling approach, while OR-weighted and CF-weighted estimates are calculated using the weighting approach. We observe that the weighted estimates have low (almost zero) variability across different runs. This is expected as the weighting approach uses all the units for calculating estimates, and biasing function used in the experiments produces deterministic probabilities. The variability in sub-sampled estimates comes from sub-sampling different samples of data in every run. Overall, we see that the estimates for weighting estimates are in the range of the estimates obtained by sub-sampling, suggesting an overall equivalence of the approaches.

\subsection{Correlation Analysis}
\begin{figure*}
\begin{tabular}{cc}
\includegraphics[width=.5\textwidth]{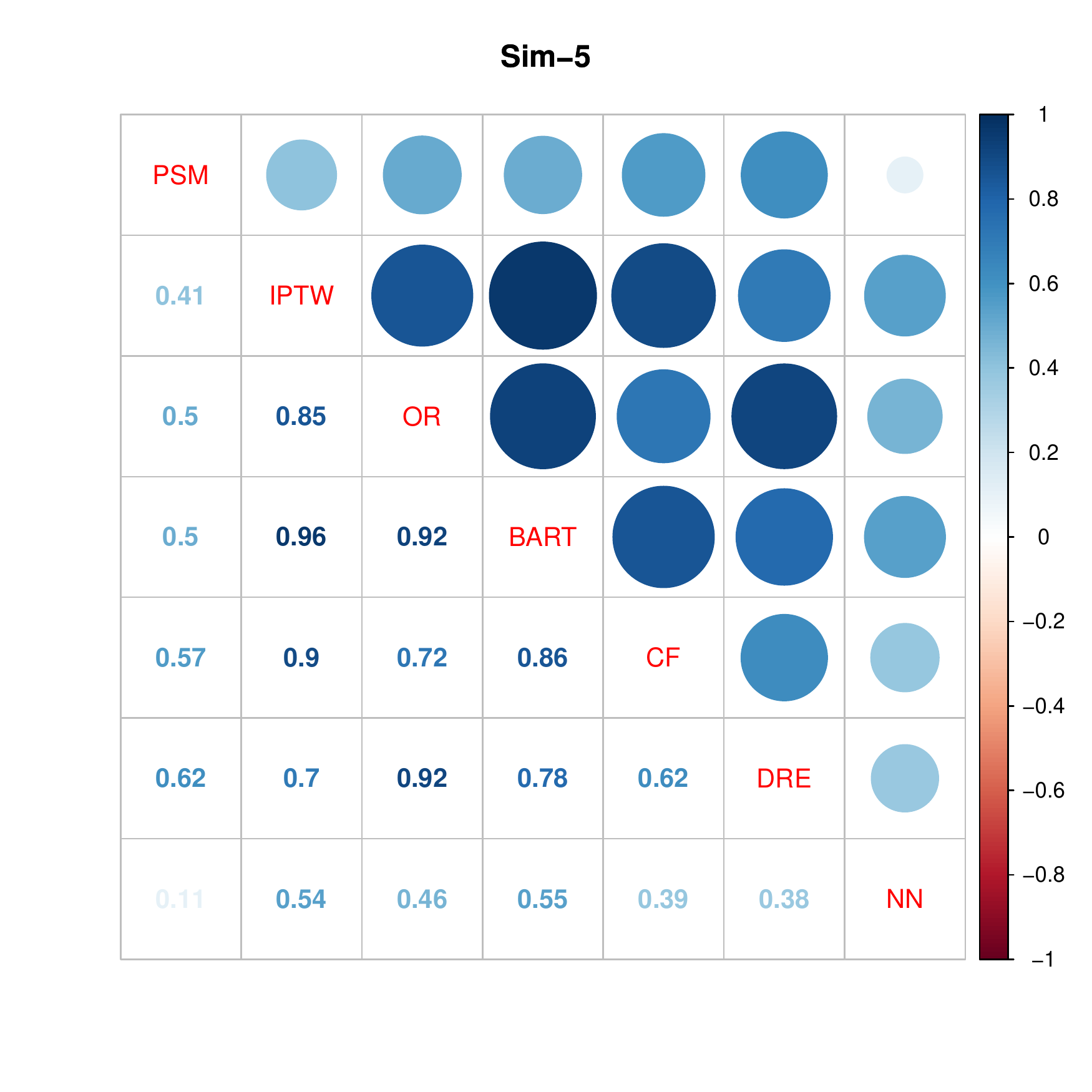} & \includegraphics[width=.5\textwidth]{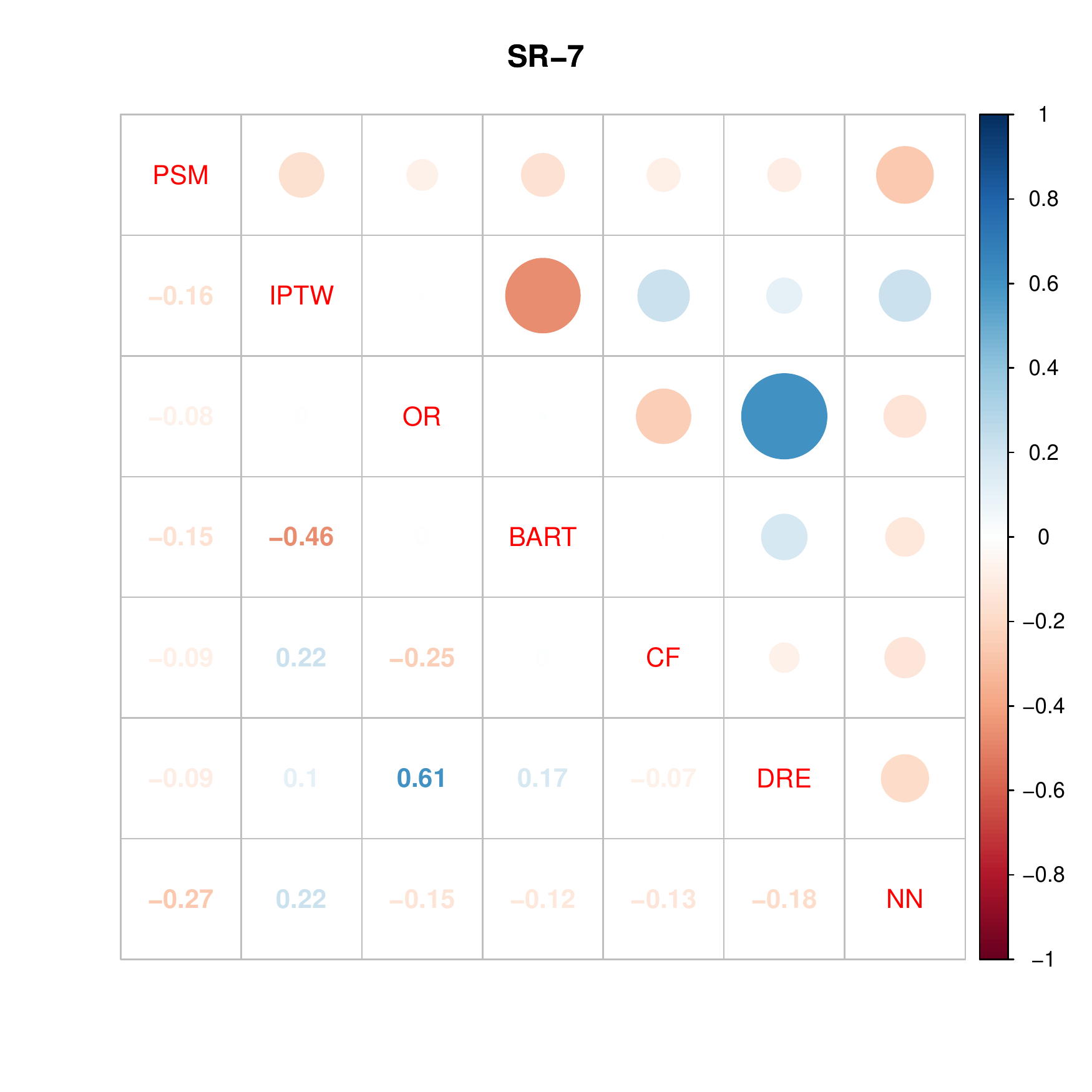} \\
\includegraphics[width=.5\textwidth]{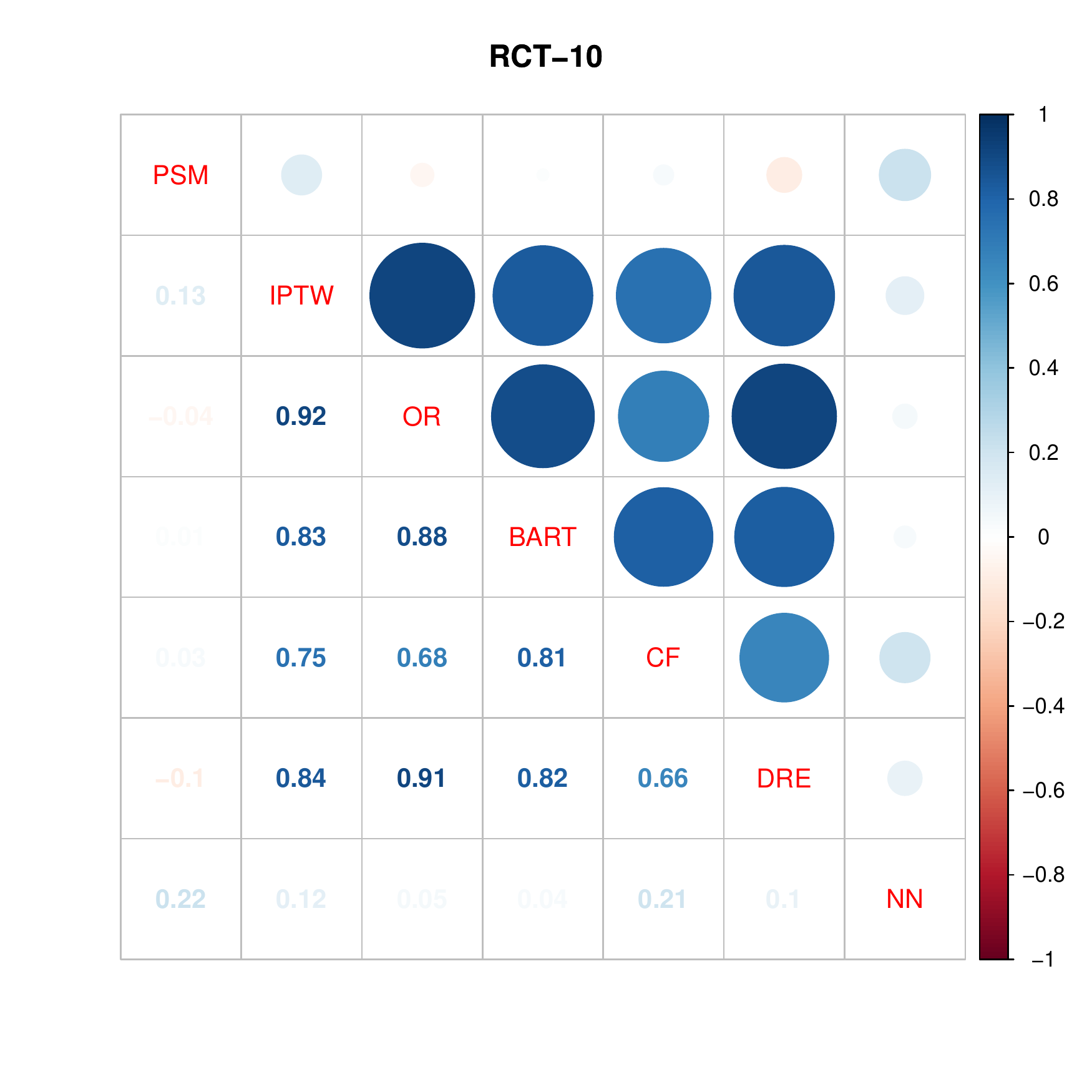} & \includegraphics[width=.5\textwidth]{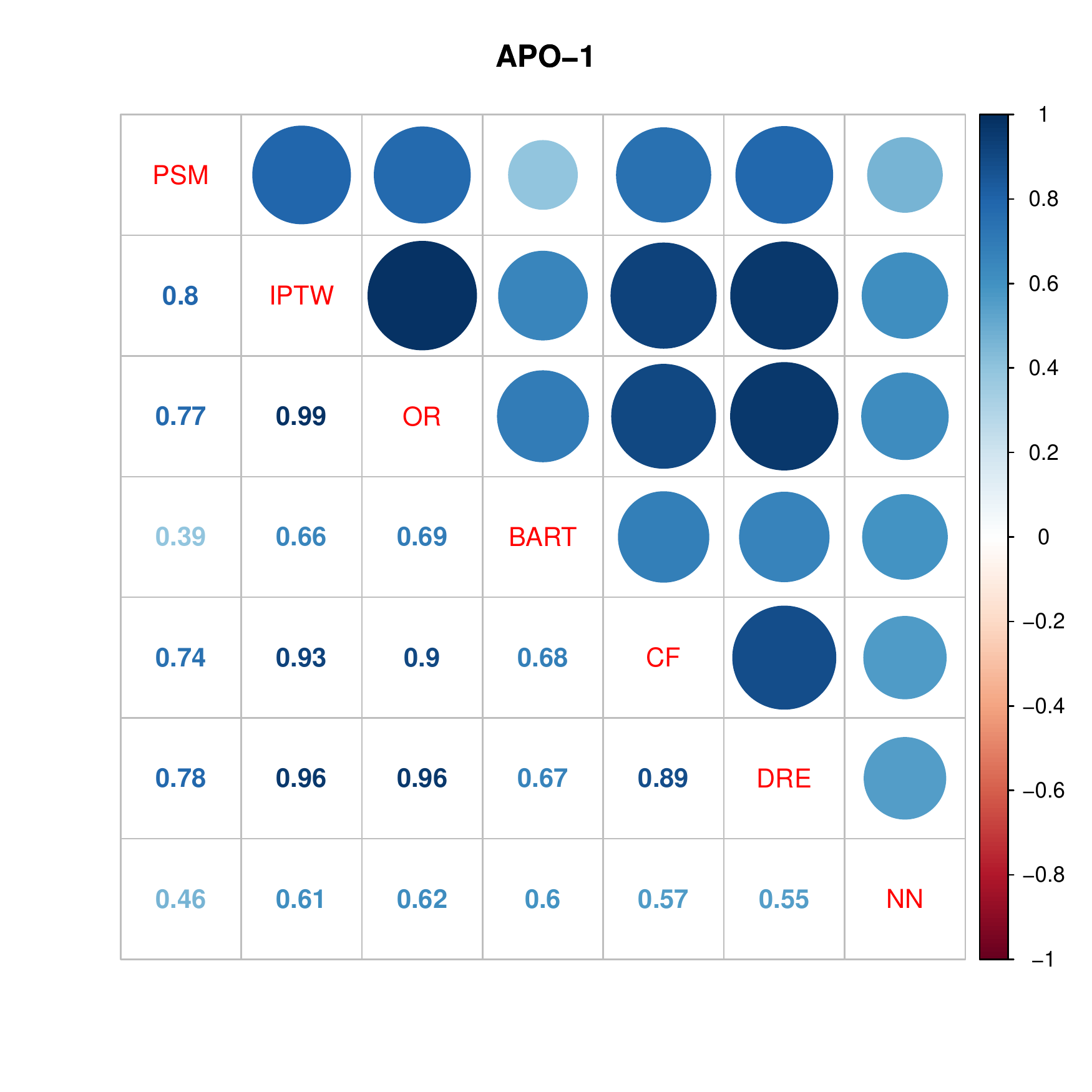} \\
\end{tabular}
\caption{Correlation matrices for four data sets.  In most cases, error is highly correlated}
\label{fig:correlation}
\end{figure*}

While the ranges of variability for most methods are the same, this doesn't guarantee that each method is producing the same result for each of the 30 trials. Error for each method could be uncorrelated with the others, suggesting that an ensemble approach might improve performance.  To test this, we computed a correlation matrix for each data set, calculating correlation across the 30 trials for each method.  Results for a few representative data sets are shown in Figure \ref{fig:correlation}.  In most cases, the correlation is the weakest with the neural network method, and is generally weaker with propensity score matching.  For all other methods, though, errors are highly correlated.  There are some exceptions, as in SR-7.  The reason for these varies.  In the case of SR-7, this is likely a result of the low variability across the 30 trials.

\subsection{Overall Mean}
\begin{figure*}
\begin{minipage}[b]{.45\textwidth}
\centering
\includegraphics[width=\textwidth]{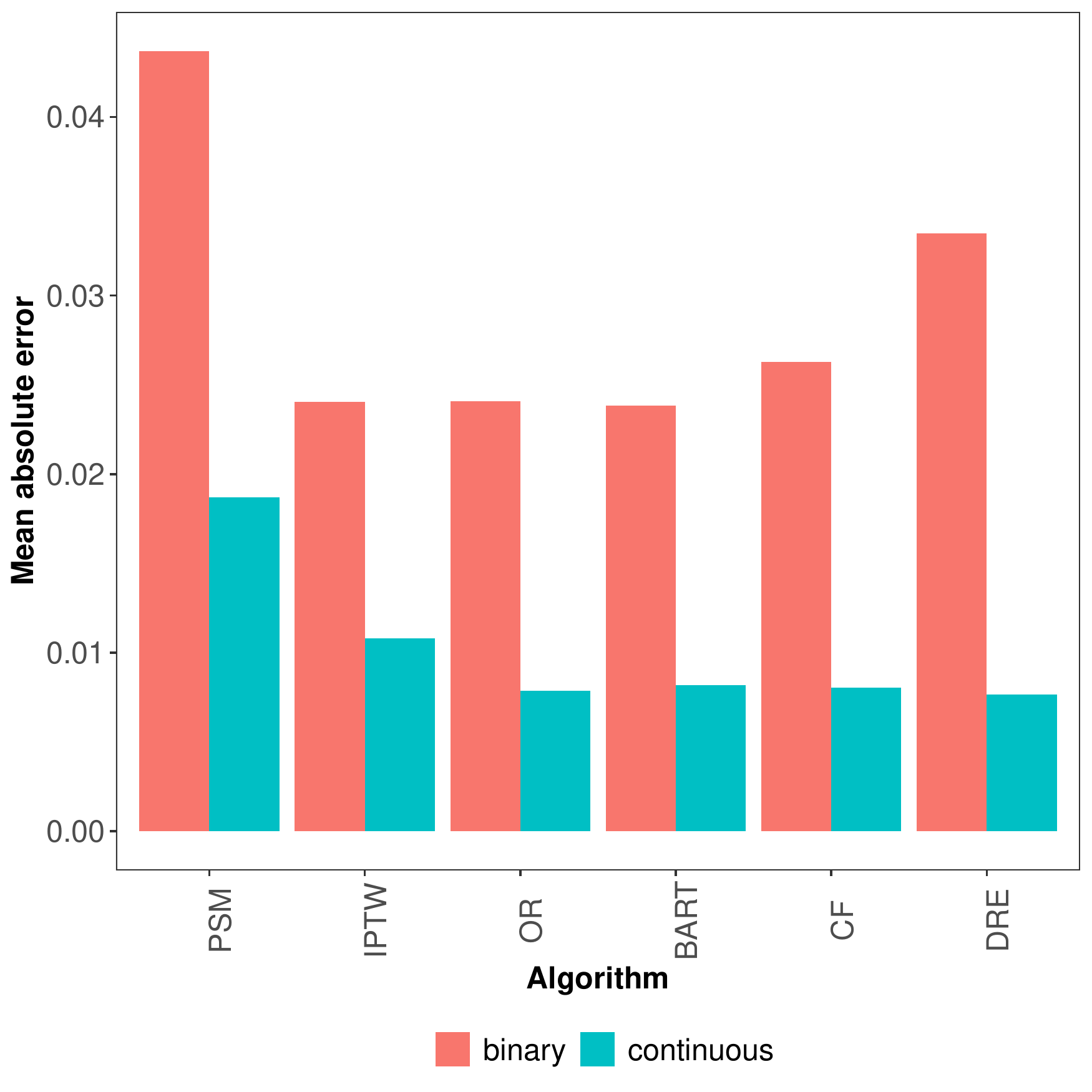}
\caption{Overall mean absolute error by algorithm}
\label{fig:overall_mean}
\end{minipage}
\hfill
\begin{minipage}[b]{.45\textwidth}
\centering
\includegraphics[width=\textwidth]{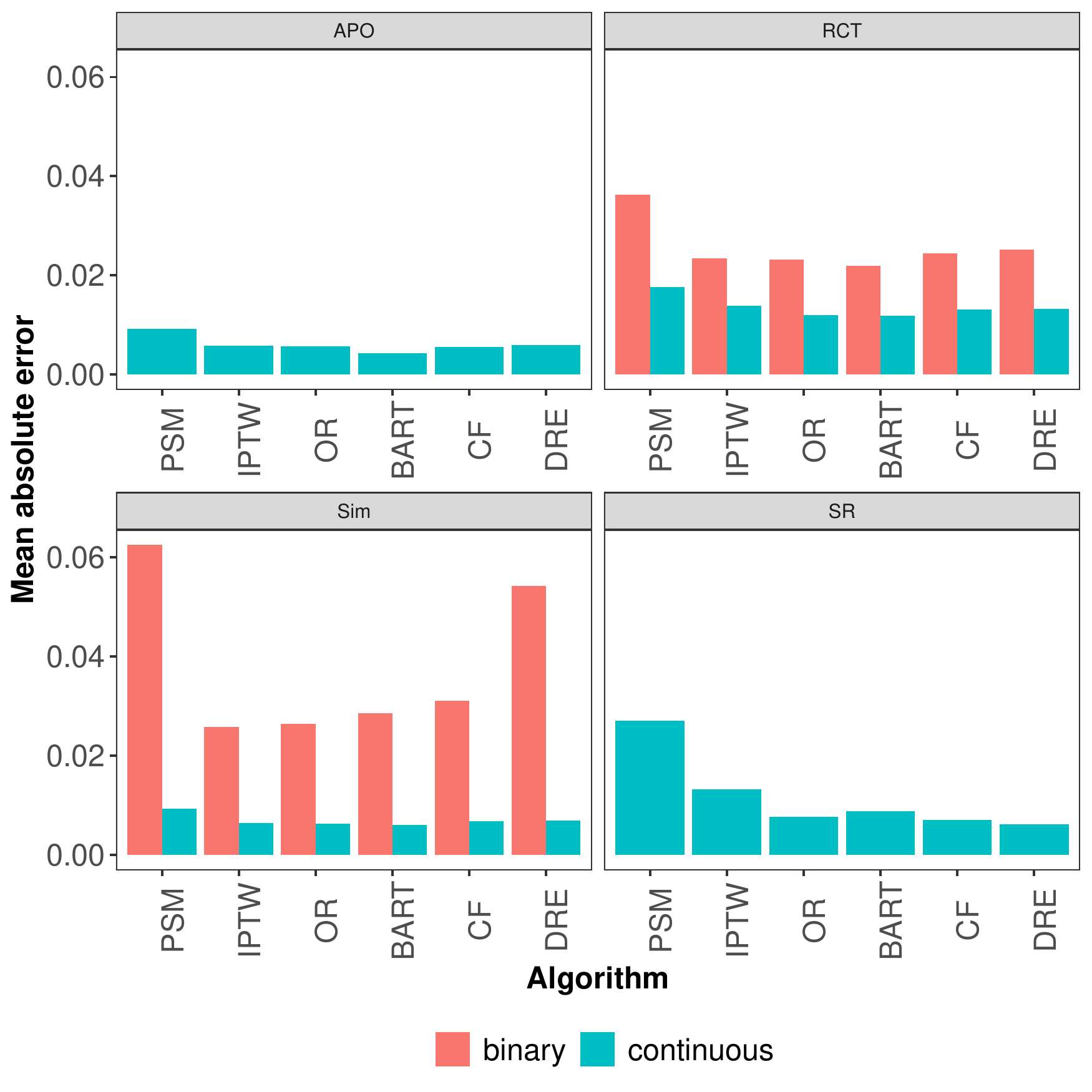}
\caption{Overall mean absolute error by algorithm, by source of data}
\label{fig:mean_by_data_source}
\end{minipage}
\end{figure*}

Figure \ref{fig:overall_mean} shows overall mean performance for each algorithm.  As observed above, propensity score matching has the highest error overall.  In addition, doubly-robust estimation appears to have higher error for data sets with binary outcomes.  More nuance can be seen in Figure \ref{fig:mean_by_data_source}, which shows mean error by data source.  The higher error for doubly robust estimation appears to be primarily for simulator data sets.  For the other data sources, mean performance is fairly consistent across algorithms.

\bibliography{references.bib}